\documentclass[12pt]{article}
\usepackage{amsmath}
\usepackage{amssymb}
\usepackage{amscd}
\usepackage{pstricks}
\usepackage{pst-all}
\usepackage[all]{xy}
    \SelectTips{cm}{}
\usepackage[mathscr]{euscript}
\usepackage{graphicx}
\usepackage{booktabs}
\usepackage{theorem}
\theoremstyle{plain}
\newtheorem{thm}{Theorem}[section]
\newtheorem{prp}[thm]{Proposition}

\newcommand{\qed}{\hbox{\rule[-2pt]{3pt}{6pt}}}

\usepackage{cite}

\makeatletter
 \@addtoreset{equation}{section}
 
 \makeatother

\begin{document}

\title{Topos Quantum Theory on Quantization-Induced Sheaves}
\author{Kunji Nakayama\footnote{e-mail: nakayama@law.ryukoku.ac.jp}
\\
Faculty of Law\\
Ryukoku University\\
Fushimi-ku, Kyoto 612-8577}

\maketitle

\def\Rmath{{\mathbb{R}}}

\def\Set{{\bf Set}}
\def\Sets{{\bf Sets}}
\def\Sh{{\rm Sh}}
\def\op{{\rm op}}

\def\Ocal{{\cal O}}
\def\Ocalop{{\cal O}^{\rm op}}
\def\OcalBorelR{\Ocal \times \mathfrak{B}_{\mathbb{R}}}
\def\Obj{{\rm Obj}}
\def\Mor{{\rm Mor}}

\def\Ahat{\hat{A}}
\def\Bhat{\hat{B}}
\def\Ehat{\hat{E}}
\def\Hhat{\hat{H}}
\def\Rhat{\hat{R}}
\def\hhat{\hat{h}}
\def\khat{\hat{k}}
\def\lhat{\hat{l}}
\def\gammahat{\hat{\gamma}}

\def\Aboldhat{\widehat {\bf A}}
\def\Eboldhat{\widehat {\bf E}}
\def\Cboldhat{\widehat {\bf C}}
\def\Pboldhat{\widehat {\bf P}}
\def\Vboldhat{\widehat {\bf V}}
\def\Vcalhat{\widehat {\cal V}}

\def\Aboldop{{\bf A^{{\rm op}}}}
\def\Cboldop{{\bf C^{{\rm op}}}}
\def\Eboldop{{\bf E^{{\rm op}}}}
\def\Vboldop{{\bf V^{{\rm op}}}}
\def\Vcalop{{\cal V^{{\rm op}}}}

\def\Aboldop{{\bf A}^{\rm op}}

\def\Vboldupsilon{{\bf V}_{\uptilde}}
\def\Vboldupsilonop{{\bf V}_{\uptilde}^{\rm op}}
\def\Vboldupsilonhat{\hat{{\bf V}_{\uptilde}}}

\def\CboldOcal{{\bf C}_{{\cal O}}}
\def\CboldOcalop{{\bf C}_{{\cal O}}^{\rm op}}
\def\CboldOcalhat{\widehat{{\bf C}_{{\cal O}}}}

\def\alphabold{{\bf \alpha}}

\def\ybold{{\bf y}}
\def\Abold{{\bf A}}

\def\Bbold{{\bf B}}
\def\Cbold{{\bf C}}
\def\Gbold{{\bf G}}
\def\Ebold{{\bf E}}
\def\Fbold{{\bf F}}
\def\Mbold{{\bf M}}
\def\Nbold{{\bf N}}
\def\Kbold{{\bf K}}
\def\Hbold{{\bf H}}
\def\Lbold{{\bf L}}
\def\Pbold{{\bf P}}
\def\Vbold{{\bf V}}
\def\Sbold{{\bf S}}
\def\Tbold{{\bf T}}
\def\Omegabold{{\bf \Omega}}

\def\Kcal{{\cal K}}

\def\Pcal{{\cal P}}
\def\Qcal{{\cal Q}}
\def\betatilde{\tilde{\beta}}

\def\Subsets{{\rm Subsets}}

\def\Hom{{\rm Hom}}
\def\Sub{{\rm Sub}}
\def\Hyp{{\rm Hyp}}
\def\dom{{\rm dom}}
\def\cod{{\rm cod}}
\def\Hyp{{\rm Hyp}}

\def\Flt{{\cal F}}
\def\Dcal{{\mathfrak D}}
\def\Ecal{{\mathfrak E}}
\def\Fcal{{\mathfrak F}}
\def\Lcal{{\mathfrak L}}
\def\Ocal{{\cal O}}
\def\Scal{{\cal S}}

\def\Amath{{\mathfrak A}}
\def\Bmath{{\mathfrak B}}
\def\Cmath{{\mathfrak C}}
\def\Emath{{\mathfrak E}}

\def\Scalbar{{\bar{\cal S}}}
\def\Hcal{{\cal H}}
\def\Ccal{{\cal C}}
\def\Gcal{{\cal G}}
\def\Rcal{{\cal R}}

\def\Tcal{{\cal T}}
\def\Vcal{{\cal V}}
\def\tmath{{\mathfrak{t}}}
\def\Dmath{{\mathfrak D}}
\def\Fmath{{\mathfrak F}}
\def\Vmath{{\mathfrak V}}

\def\CcalR{{{\cal C}_R}}
\def\CcalRe{{{\cal C}_R^{e \downarrow}}}
\def\Ocaltilde{{\cal O}/_{\sim}}

\def\ahat{\hat{a}}
\def\Fhat{\hat{F}}
\def\Phat{\hat{P}}
\def\Hhat{\hat{H}}
\def\Ohat{\hat{O}}
\def\Ihat{\hat{I}}
\def\alphahat{\hat{\alpha}}

\def\Ghat{\hat{G}}

\def\Com{{\rm Com}}
\def\ES{{\rm ES}}
\def\op{{\rm op}}
\def\Onebold{{\bf 1}}
\def\Mor{{\rm Mor}}
\def\Obj{{\rm Obj}}
\def\Bcal{{\cal B}}
\def\End{{\rm End}}

\def\deltabold{\boldsymbol{\delta}}
\def\deltaboldj{\boldsymbol{\delta}_{j}}

\def\Pmathbb{\mathbb{P}}
\def\Rmathbb{\mathbb{R}}
\def\Smathbb{\mathbb{S}}
\def\Tmathbb{\mathbb{T}}
\def\Tmathbbphi{\mathbb{T}^{| \varphi \rangle}}
\def\Tmathbbrhor{\mathbb{T}^{\rho, \, r}}

\def\varphivec{| \varphi \rangle}
\def\varphipro{| \varphi \rangle \langle \varphi |}

\def\taurhor{\tau^{\rho, \, r}}
\def\tauphi{\tau^{| \varphi \rangle}}

\def\subseteqVbold{\subseteq_{\Vbold}}

\def\true{{\rm true}}
\def\onepoint{\{ \, \cdot \,\}}
\def\id{{\rm id}}
\def\pt{{\rm pt}}
\def\dB{{\rm dB}}
\def\cl{{\rm cl}}
\def\Cl{{\cal C}l}

\def\alphabreve{\breve{\alpha}}
\def\uptilde{\tilde{\upsilon}}

\def\vertin{\mathop{{\rotatebox{90}{$\in$}}}}

\def\vertsim{\mathop{{\rotatebox{90}{$\sim$}}}}
\def\fortyfivesim{\mathop{{\rotatebox{-45}{$\sim$}}}}

\def\tr{{\rm tr}}
\def\qed{\hspace{\fill}$\square$}

\def\gammabold{\boldsymbol{\gamma}}
\def\gammaboldmax{\boldsymbol{\gamma}^{\vee}}
\def\gammaboldmin{\boldsymbol{\gamma}^{\wedge}}
\def\jmathboldmax{\boldsymbol{\jmath}^{\vee}}
\def\jmathboldmin{\boldsymbol{\jmath}^{\wedge}}
\def\jmathbold{\boldsymbol{\jmath}}
\def\imathbold{\boldsymbol{\imath}}
\def\imathboldmax{\boldsymbol{\imath}^{\vee}}
\def\imathboldmin{\boldsymbol{\imath}^{\wedge}}

\def\Tpre{{\rm Sub}_{{\rm filt}} (\Pmathbb_{\cl} \Sigma)}

\def\Ucalflat{\mathcal{U}^{\flat}}

\begin{abstract}

In this paper, we construct a sheaf-based topos quantum theory.
It is well known that a topos quantum theory can be constructed on the topos of presheaves on the category of commutative von Neumann algebras of bounded operators on a Hilbert space.
Also, it is already known that quantization naturally induces a Lawvere-Tierney topology on the presheaf topos.
We show that a topos quantum theory akin to the presheaf-based one can be constructed on sheaves defined by the quantization-induced Lawvere-Tierney topology.
That is, starting from the spectral sheaf as a state space of a given quantum system,
we construct sheaf-based expressions of physical propositions and truth objects, and thereby give a method of truth-value assignment to the propositions.
Furthermore, we clarify the relationship to the presheaf-based quantum theory.
We give translation rules between the sheaf-based ingredients and the corresponding presheaf-based ones.  
The translation rules have `coarse-graining' effects on the spaces of the presheaf-based ingredients;
a lot of different proposition presheaves, truth presheaves, and presheaf-based truth-values are translated to a proposition sheaf, a truth sheaf, and a sheaf-based truth-value, respectively.
We examine the extent of the coarse-graining made by translation.
\end{abstract}

\newpage

\section{Introduction}

Since Isham \cite{I97} applied topos theory to history quantum theory, topos
theoretic approach to quantum theory has been studied by many researchers
\cite{IB98,BI99,HBI00,BI02,DI08a,DI08b,DI08c,DI08d,HLS09,F10,HLS11,HLSW11,DI11,DI12,W13,F13a,F13b}. In this approach, quantum theory is reformulated within a framework
of intuitionistic (hence, multi-valued) logic. 
Every physical proposition about
a given quantum system is assigned a truth-value without falling foul of the
Kochen-Specker no go theorem \cite{KS67}. 
Therefore, the topos approach permits some kind of realistic interpretation regarding values of physical quantities that does not require things like the notion of measurement.
Because of this, it can provide a promising framework for quantum gravity theory and quantum cosmology.

There are a few different ways of topos approach. 
Among them, we focus on the formalism made by D\"{o}ring and Isham \cite{DI08a,DI08b,DI08c,DI08d,DI11,DI12}. 
They adopted the topos of presheaves on the category of commutative von
Neumann algebras of bounded operators on a Hilbert space.
In their theory, the spectral presheaf plays a key role similar to state space of classical physics. 
As a result of the Kochen-Specker theorem, 
we cannot assign to every physical quantity of a quantum system a sharply determined value,
which means there are no global elements of the spectral presheaf \cite{IB98,BI99,HBI00,BI02}.
In this respect, the spectral presheaf is largely different from the state space of classical physics,
since the latter consists of points, each of which corresponds to a state where every physical quantity has sharply determined value.
Nonetheless, the spectral presheaf can work as a
state space in that every physical proposition about a given quantum system
can be expressed as its subobject, as every physical proposition about a
classical system can be identified with its extensional expression, i.e., a subset
of state space. 
By regarding the spectral presheaf as state space and using it in a topos theoretical framework,
D\"{o}ring and Isham succeeded in giving a method that assigns to every physical
proposition a truth-value.

D\"{o}ring and Isham's theory is an abstract, general theory;
it does not need to be related to concrete classical systems,
like ordinary quantum theories that are axiomatically or algebraically formulated on Hilbert spaces or $C^{*}$-algebras. 
This is the case for the other topos quantum theories obtained so far.
If quantization of a classical system is taken into consideration, 
however, some extra structures are induced on the topos on
which a quantum theory is formulated.
 In fact, Nakayama \cite{N13} showed that
quantization that is given by a function from classical observables to self-adjoint
operators on a Hilbert space naturally induces a Lawvere-Tierney
topology on the presheaf topos of D\"{o}ring and Isham. 
It is well-known that any Lawvere-Tierney topology defines sheaves, 
and furthermore, the collection of all such sheaves also forms a topos \cite{MM92}. 
Thus, from the presheaf topos, we obtain another topos consisting of sheaves via quantization.

One question would arise. Can we construct a quantum theory on the
topos of quantization-induced sheaves? 
One of the purposes of the present paper is to give an affirmative answer to the question. 
We can construct a topos quantum theory on the quantization-induced sheaves in a way akin to
the presheaf-based theory of D\"{o}ring and Isham. 
Such a theory could be canonical as a theory of the system quantized from the classical one since quantum observables corresponding to classical ones are identified therein.

Furthermore, the theory on quantization-induced
sheaves can be formulated by means of topos-theoretic ingredients smaller than those of the presheaf-based theory. 
For example, as we will see, the space of truth-values
of the quantization-induced topos is smaller than that of the matrical
topos of presheaves. 
This is because, for each sheaf-based truth-value,
there are a lot of different presheaf-based ones that can be regarded as its
`translations', and conversely, a lot of different presheaf-based truth-values
are translated to one and the same sheaf-based one. 
The same holds for the space of propositions and that of truth objects, 
because each sheaf-based proposition and each
truth object have a lot of different presheaf-based translations. 
We call these properties coarse-graining made by translation.

Another question would arise. 
To what degree do the spaces of
presheaf-based truth-values, propositions, and truth objects get coarse-grained
via translation? 
In this paper, we answer this
question to some extent. 
We give translation rules between the sheaf-based
ingredients and the presheaf-based ones, and for an arbitrarily given sheaf-based
one, 
we explicitly construct corresponding subspaces consisting of its
presheaf-based translations that are regarded as the same from the sheaf-based viewpoint.

The present paper is organized as follows. 
In section \ref{sec:review of Nakayama 2013}, we briefly review
Nakayama's result \cite{N13} about quantization-induced topologies and sheaves.
Further, additional explanation about some related notions that we will need
in later sections are given. 
In section \ref{sec:truth-value valuation}, we develop the sheaf-based method
of truth-value assignment.
This is done along the line of the presheaf-based method given by D\"{o}ring and Isham \cite{DI12},
which we briefly summarize in appendix \ref{sec:DI} for referential convenience. 
(We should, however, note that the main purpose of D\"{o}ring and Isham \cite{DI12}
is not to give the method itself but to propose a new interpretation for
quantum probabilities, which is beyond the scope of the present paper.) 
In section \ref{sec:translation rule}, 
we give rules of translation of the ingredients necessary for truth-value assignment between
the sheaf-based and the presheaf-based cases. 
In section \ref{sec:coarse-graining}, we deal with
the coarse-graining problem mentioned above. 
Main results obtained therein are presented by theorems \ref{thm:coarse-graining of truth-value},
 \ref{thm:coarse-graining of proposition}, and  \ref{thm:coarse-graining of truth object}.


\section{Topos of Sheaves Induced by Quantization}
\label{sec:review of Nakayama 2013}

In this section, we give a brief review of the results by Nakayama \cite{N13}
and some supplementary explanations.
Nakayama \cite{N13} defines quantization as an injective map $\upsilon$ 
from a Lie algebra $\Ocal$, a model of classical observables \cite{I84},
to self-adjoint operators on a Hilbert space $\Hcal$.
The quantization map naturally defines a functor $\phi$ 
from the category $\Cbold(\Ocal)$ of sets of commutative classical observables
to the category $\Vbold$ of commutative von Neumann algebras of bounded operators on $\Hcal$.
The functor $\phi$ assigns to each $C \in \Cbold(\Ocal)$ 
the least commutative von Neumann algebra that includes ${\rm e}^{{\rm i} \upsilon(C)}$. 
We define a functor $\psi:\Vbold \to \Cbold(\Ocal)$ by
\begin{equation}
\psi(V) := \{ a \in \Ocal \;|\; {\rm e}^{{\rm i} \upsilon(a)} \in V \}.
\end{equation}
The functors $\phi$ and $\psi$ give a Galois connection between $\Cbold(\Ocal)$ and $\Vbold$.

The endofunctor $\flat := \phi \psi :  \Vbold \to \Vbold$ induces a Grothendieck topology $J$ on $\Vbold$,
which is defined by for each $V \in \Vbold$,
\begin{equation}
J(V) := \{ \omega  \in \Omega(V)\;|\; \flat(V) \in \omega\},
\label{eq:Grothendieck}
\end{equation}
where $\Omega$ is the subobject classifier of the topos $\Vboldhat \equiv \Set^{\Vbold^{\op}}$ of presheaves on $\Vbold$.

As is well-known, every Grothendieck topology on $\Vbold$ is equivalent to a Lawvere-Tierney topology on $\Vboldhat$. (As for general theory of topoi, see e.g., MacLane and Moerdijk's textbook \cite{MM92}.)
The Grothendieck topology (\ref{eq:Grothendieck}) gives the Lawvere-Tierney topology $\Omega \xrightarrow{j} \Omega$ defined by, for each $V \in \Vbold$ and $\omega \in \Omega(V)$,
\begin{equation}
j_{V}(\omega):= \{ V' \subseteqVbold V \;|\; \flat(V') \in \omega \},
\label{eq:LT}
\end{equation}
where $V' \subseteqVbold V$ means that $V'$, $V \in \Vbold$ and $V' \subseteq V$.

Each Lawvere-Tierney  topology is equivalent to a closure operator.
In the present case given by (\ref{eq:LT}), for each presheaf $Q \in \Vboldhat$
and its subobject $S \in \Sub(Q)$,
the closure $\bar{S}$ of $S$ in $Q$ is defined by
\begin{equation}
\bar{S}(V) := \{ q \in Q(V) \;|\; Q(\flat(V) \hookrightarrow V)(q) \in S(\flat(V))  \}.
\label{eq:closure}
\end{equation}

Any Lawvere-Tierney topology $j$ on $\Vboldhat$ defines sheaves as follows:
Let $S \in \Sub(Q)$ be dense in $Q$,
that is, $\bar{S} = Q$.
Then, a presheaf $R$ is called a sheaf 
associated with a topology $j$, or simply, $j$-sheaf,
if and only if, for any morphism $\lambda \in \Hom(S,R)$,
there exists one and only one morphism $\mu \in \Hom(Q,R)$
that makes the diagram
\begin{equation}
\xymatrix{
S \ar [rr] ^{\lambda} \ar @{>->} [dd] _{\mbox{dense}} && R \\
&& \\
Q \ar [rruu] _{\mu} &&  
}
\label{eq:definition of sheaf}
\end{equation}  
 commute.
All $j$-sheaves and all morphisms between them form a topos,
which is denoted by $\Sh_{j} \Vboldhat$.

Sheaves associated with the topology (\ref{eq:LT}) are expressed by
the functor $\flat^{*}:\Vboldhat \to \Vboldhat$ that is defined by,
for each $Q \in \Vboldhat$,
\begin{equation}
(\flat^{*}Q)(V) := Q(\flat(V)) ,
\end{equation}
and for any $V' \subseteqVbold V$,
\begin{equation}
(\flat^{*}Q)(V' \hookrightarrow V) := Q(\flat(V') \hookrightarrow \flat(V)). 
\end{equation}
We can show that a presheaf $Q$ is a $j$-sheaf if and only if $Q$ is isomorphic to $\flat^{*}Q$.
To make the condition more precise, we define a morphism $Q \xrightarrow{\zeta_{Q}} \flat^{*} Q$ by
%
%
$(\zeta_{Q})_{V} := Q(\flat(V) \hookrightarrow V)$.
Then, $Q$ is a $j$-sheaf if and only if $\zeta_{Q}$ is isomorphic.
We should note that $\zeta_{Q}$ is natural with respect to $Q \in \Vboldhat$.
That is, $\zeta$ is a natural transformation from the identity functor $I:\Vboldhat \to \Vboldhat$ to the functor $\flat^{*}:\Vboldhat \to \Vboldhat$.
Furthermore, we should note that $\flat^{*}$ is in fact an associated sheaf functor (a sheafification functor) from $\Vboldhat$ to $\Sh_{j} \Vboldhat$.

Returning to the diagram (\ref{eq:definition of sheaf}), 
we note that the morphism $\mu$ is given by 
\begin{equation}
\mu = \zeta_{R}^{-1} \circ \flat^{*} \lambda \circ \zeta_{Q},
\label{eq:sheaf2}
\end{equation}
since the naturality of $\zeta$ makes the diagram
\begin{equation}
\xymatrix{
S \ar [rrrr] ^{\lambda} \ar @{>->}_{\rm{dense}} [rrd] \ar [ddd] _{\zeta_{S}} &&&& R \ar [ddd] ^{\zeta_{R}} _{\vertsim} \\
&& Q \ar [rru] _{\mu} \ar [d] ^{\zeta_{Q}} && \\
&& \flat^{*} Q \ar [rrd] ^{\flat^{*}\mu = \flat^{*}\lambda} && \\
\flat^{*}S \ar @{->} [rrrr] _{\flat^{*}\lambda} \ar @{=} [rru] &&&& \flat^{*}R \\
}
\end{equation}
commute. 
Here, this digram reflects the fact that $\flat^{*}S = \flat^{*} \bar{S } = \flat^{*} Q$.

In our formalism, truth-values of physical propositions are taken on the subobject classifier $\Omega_{j}$ of $\Sh_{j} \Vboldhat$.
That is, they are given as global elements 
$\begin{xy} {\ar @{>->}  (0,0)*{1_{ }\;};(8,0)*{\;\Omega_{j}}}  \end{xy}$ $\in \Gamma \Omega_{j}$ of $\Omega_{j}$.
As is well-known, $\Omega_{j}$ is the equalizer of $\Omega \xrightarrow{1_{\Omega}} \Omega$ and
$\Omega \xrightarrow{j} \Omega$.
In the present case, $\Omega_{j}$ is a subobject of $\Omega$ given by
\begin{equation}
\Omega_{j}(V) := \{ \omega \in \Omega(V) \;|\; \forall V' \subseteqVbold V \;
( \flat(V') \in \omega \Rightarrow V' \in \omega  )\}.
\label{eq:subobject classifier of quantization topos}
\end{equation}
Since for each $V \in \Vbold$, $\Omega_{j}(V)$ contains 
the set  $\tmath_{V}$ of all subalgebras of $V$ as the top element,
the truth arrow $\true_{j} \in \Gamma \Omega_{j}$ is given by
\begin{equation}
(\true_{j})_{V}:= \tmath_{V}.
\end{equation}

Later, we will deal with power objects in $\Sh_{j} \Vboldhat$.
As is well-known, 
the power object $\Pmathbb_{j} R \equiv \Omega_{j}^{R}$ of 
a j-sheaf $R$ can be calculated in $\Vboldhat$.
That is, 
for each $V \in \Vbold$,
\begin{eqnarray}
(\Pmathbb_{j} R)(V) 
&=& 
\Hom(R_{\downarrow V}, (\Omega_{j})_{\downarrow V}) \nonumber\\
\label{eq:power object 1a}
& \simeq &
\Hom(R_{\downarrow V}, \Omega_{j}),
\label{eq:power object 1b}
\end{eqnarray}
where $R_{\downarrow V}$ and $ (\Omega_{j})_{\downarrow V}$ are downward restrictions as presheaves,
the definition of which is given by (\ref{eq:downarrow 1}) and (\ref{eq:downarrow 2}).
(Since $\Sh_{j}\Vboldhat$ is a full subcategory of $\Vboldhat$, $\Hom_{\Sh_{j}\Vboldhat}(A,B) = \Hom_{\Vboldhat}(A,B)$ for arbitrary sheaves $A$ and $B$. 
So we simply write $\Hom(A,B)$ for both of them omitting the subscripts $\Sh_{j}\Vboldhat$ and $\Vboldhat$.)
Also, for $V' \subseteqVbold V$ and $\lambda \in (\Pmathbb_{j} R)(V)$,
$\lambda |_{V'} \equiv (\Pmathbb_{j}R)(V' \hookrightarrow V)(\lambda) $ 
is defined as the morphism that makes the diagram
\begin{equation}
\xymatrix{
R_{\downarrow V'} \ar [rr] ^{\lambda|_{V'}} \ar @{>->} [dd] && \Omega_{j} \\
&& \\
R_{\downarrow V} \ar [rruu] _{\lambda} && \\
}
\label{eq:power object 2}
\end{equation}
commute.

In order to give another, more useful expression of the power object $\Pmathbb_{j} R$,
we note that it is a sheaf representing the collection $\Sub_{j} (R)$ of all subsheaves of $R$.
Let $Q$ be a presheaf.
As we will see below, $\Pmathbb_{j}(\flat^{*} Q)$ can be expressed as
\begin{equation}
\Pmathbb_{j}( \flat^{*} Q)(V) \simeq \Sub_{j} (\flat^{*} (Q_{\downarrow V}))
\label{eq:power object' 1}
\end{equation}
and
\begin{equation}
\Pmathbb_{j} (\flat^{*} Q)(V' \hookrightarrow V) : \Sub_{j} (\flat^{*} (Q_{\downarrow V})) \to \Sub_{j} (\flat^{*} (Q_{\downarrow V'}));
S \mapsto \flat^{*}(S_{\downarrow V'}).
\label{eq:power object' 2}
\end{equation}
In particular, since any $j$-sheaf $R$ satisfies $R \simeq \flat^{*} R$,
we have
\begin{equation}
\Pmathbb_{j} R \simeq \Pmathbb_{j}(\flat^{*}R) \simeq \Sub_{j}(\flat^{*}(R_{\downarrow V})).
\end{equation}
Expression (\ref{eq:power object' 1}) comes from the fact that 
\begin{eqnarray}
\Pmathbb_{j}(\flat^{*} Q) (V) 
& \simeq &
\Hom ((\flat^{*} Q)_{\downarrow V},\Omega_{j}) \nonumber\\
& \simeq &
\Hom (\flat^{*} (Q_{\downarrow V}),\Omega_{j}) \nonumber\\
& \simeq &
\Sub_{j}(\flat^{*} (Q_{\downarrow V})) .
\label{eq:power bijection} 
\end{eqnarray}
Here, the bijectivity between the first and second lines on (\ref{eq:power bijection}) is verified 
from the commutative diagram
\begin{equation}
\xymatrix{
(\flat^{*} Q)_{\downarrow V'} \ar [rd] ^{\kappa|_{V'}} \ar @{>->} [dd] _{\mbox{dense}} \ar @{>->} [rr]&& (\flat^{*} Q)_{\downarrow V}  \ar [ld] _{\kappa} \ar @{>->} [dd] ^{\mbox{dense}} \\
& \Omega_{j} & \\
\flat^{*} (Q_{\downarrow V'} )  \ar [ru] _{\chi|_{V'}}   \ar @{>->} [rr] && \flat^{*} (Q_{\downarrow V} ) \ar [lu] ^{\chi} \; .\\
}
\label{eq:P<=>S-relation}
\end{equation}
That is, since $\Omega_{j}$ is a $j$-sheaf, and since, as easily shown, 
$(\flat^{*} Q)_{\downarrow V}$ is dense in $\flat^{*} (Q_{\downarrow V} ) $,
 $\chi$ is uniquely determined by use of (\ref{eq:sheaf2}) for each morphism $\kappa$.

To see consistency between  (\ref{eq:power object 2}) and (\ref{eq:power object' 2}),
let $S^{\chi}$ be a subsheaf of $\flat^{*}(Q_{\downarrow V})$ corresponding to the characteristic morphism $\chi$.
Then, in the diagram
\begin{equation}
\xymatrix{
\flat^{*}(S^{\chi}_{\downarrow V'})\ar @{>->} [rr] \ar @{>->} [ddd] \ar [rd] _{!}  && S^{\chi} \ar [ld] ^{!} \ar @{>->} [ddd]  \\
& 1 \ar @{>->} [d] ^{\true_{j}} & \\
& \Omega_{j} & \\
\flat^{*}(Q_{\downarrow V'}) \ar @{>->} [rr] \ar [ru] ^{\chi|_{V'}} && \flat^{*}(Q_{\downarrow V}) \ar  [lu] _{\chi} , \\
}
\end{equation}
the trapezoid at the right hand side is a pullback,
and so is the outer square as easily shown.
Thus, also the trapezoid at the left hand side is a pullback,
which means that $\flat^{*}(S^{\chi}_{\downarrow V'}) $ is identified as the subsheaf of $\flat^{*}(Q_{\downarrow V'}) $ that corresponds to the characteristic morphism $\chi|_{V'} \equiv \Pmathbb_{j}(\flat^{*} Q)(V' \hookrightarrow V)(\chi)$.


\section{Truth-Value Assignment on Quantization-Induced Sheaves}
\label{sec:truth-value valuation}

In the theory of D\"{o}ring and Isham \cite{DI08a,DI08b,DI08c,DI08d,DI11,DI12}, 
the spectral presheaf $\Sigma$, the definition of which is given in appendix \ref{sec:DI}, plays a role of state space of a given quantum system.
Every physical proposition $P$ is assumed to be representable as a clopen subobject of $\Sigma$,
that is, an element of the collection $\Sub_{{\rm cl}}(\Sigma)$ of all clopen subobjects of $\Sigma$.
For instance, D\"{o}ring and Isham showed that each projection operator $\Phat$, which corresponds to some physical propositions in ordinary quantum theory, naturally defines a clopen subobject $\delta(\Phat)$ of $\Sigma$ via the `daseinization operator' $\delta$.
If we are given a quantum state,
we can specify propositions regarded as true.
They are represented by a truth object $\Tmathbb$, of which global elements give the truth propositions. 
If we have $\Tmathbb$, we can assign to every proposition $P$ a truth value via topos-theoretical setting.
In appendix \ref{sec:DI}, we give a brief explanation of the method of truth-value assignment developed by D\"{o}ring and Isham \cite{DI12}, the style of which is helpful for us to construct a sheaf-based theory.
(It should be emphasized, however, that the main purpose of \cite{DI12} is not to give the valuation method summarize in appendix \ref{sec:DI},
but to propose a new interpretation of quantum probabilities based on intuitionistic logic,
which is beyond the scope of the purpose of the present paper.)

In our formalism, we appropriate the `spectral sheaf' $\flat^{*}\Sigma$ for the role of state space.
Namely, every proposition is assumed to be representable as a clopen subsheaf of $\flat^{*} \Sigma$.
We, thus, regard $\Sub_{j \, \cl}(\flat^{*} \Sigma)$,
the collection of all clopen subsheaves of $\flat^{*} \Sigma$, as a proposition space.
It can be internalized to $\Sh_{j}\Vboldhat$ as a subsheaf $\Pmathbb_{j\, \cl}(\flat^{*} \Sigma)$ of $\Pmathbb_{j}(\flat^{*} \Sigma)$
that is defined by
\begin{equation}
(\Pmathbb_{j\, \cl}(\flat^{*} \Sigma))(V)
:=
\Sub_{j \, \cl}(\flat^{*}(\Sigma_{\downarrow V})) .
\label{eq:Pmathbbjcl flat Sigma}
\end{equation}
This definition really gives a presheaf because 
$ (\Pmathbb_{j \,\cl}(\flat^{*} \Sigma))(V' \hookrightarrow V)$,
i.e., the restriction of $ (\Pmathbb_{j}(\flat^{*} \Sigma))(V' \hookrightarrow V)$
to $\Sub_{j \, \cl}(\flat^{*}(\Sigma_{\downarrow V})) $,
takes values on $\Sub_{j \, \cl}(\flat^{*}(\Sigma_{\downarrow V'}))$. 
Furthermore,
\begin{prp}
The presheaf $\Pmathbb_{j \,\cl}(\flat^{*} \Sigma)$ is a $j$-sheaf.
\end{prp}
Proof.
We have
\begin{eqnarray}
\flat^{*}(\Pmathbb_{j\, \cl}(\flat^{*} \Sigma))(V)
& = &
(\Pmathbb_{j\, \cl}(\flat^{*} \Sigma))(\flat (V)) \nonumber\\
& = &
\Sub_{j \, \cl}(\flat^{*}(\Sigma_{\downarrow \flat(V)}))  \nonumber\\
& = &
\Sub_{j \, \cl}(\flat^{*}(\Sigma_{\downarrow V}))  \nonumber\\
& = &
(\Pmathbb_{j\, \cl}(\flat^{*} \Sigma))(V),
\end{eqnarray}
where from the second line to the third, we used  (\ref{eq:flat1}).
Furthermore, for each $S \in \Sub_{j \, \cl}(\flat^{*}(\Sigma_{\downarrow V}))$,
\begin{eqnarray}
(\zeta_{\Pmathbb_{j \, \cl}(\flat^{*}\Sigma)})_{V}(S)
& = &
\flat^{*}(S_{\downarrow \flat(V)}) \nonumber\\
& = &
\flat^{*}(S_{\downarrow V}) \nonumber\\
& = &
S.
\end{eqnarray}
Therefore, $\zeta_{\Pmathbb_{j \, \cl}(\flat^{*}\Sigma)}$ is a natural isomorphism.
\qed

As is well-known, $\Sub_{j}(\flat^{*} \Sigma) \simeq \Gamma(\Pmathbb_{j}(\flat^{*} \Sigma)) :=\Hom(1, \Pmathbb_{j} (\flat^{*} \Sigma))$. 
That is, every $S \in \Sub_{j}(\flat^{*} \Sigma)$ has its name $\lceil S \rceil_{j} \in \Gamma(\Pmathbb_{j} (\flat^{*} \Sigma))$
defined by
\begin{equation}
(\lceil S \rceil_{j})_{V} := \flat^{*}(S_{\downarrow V}),
\end{equation}
and every $s \in \Gamma(\Pmathbb_{j}(\flat^{*}\Sigma))$ has its inverse, i.e., the subsheaf $\lceil s \rceil_{j}^{-1}$ of $\flat^{*} \Sigma$ given by
\begin{equation}
\lceil s \rceil_{j}^{-1}(V) := (s_{V})(V).
\end{equation}
It is obvious that, for any $S \in \Sub_{j}(\flat^{*}(\Sigma_{\downarrow V})) $, 
$\lceil S \rceil_{j} \in \Gamma (\Pmathbb_{j \,\cl}(\flat^{*} \Sigma))$
if and only if $S$ is a proposition, i.e., $S \in \Sub_{j \, \cl}(\flat^{*}(\Sigma_{\downarrow V})) $.
Furthermore, for each proposition $P\in \Sub_{j \, \cl}(\flat^{*}\Sigma)$, 
the diagram
\begin{equation}
\xymatrix{
1\ar @{>->} [rr] ^(0.4){\lceil P \rceil_{j}} \ar @{>->} [rrdd] _{\lceil P \rceil_{j}}    && \Pmathbb_{j \, \cl}(\flat^{*}\Sigma) \ar @{>->} [dd] \\
&& \\
&& \Pmathbb_{j}(\flat^{*}\Sigma)  \\
}
\end{equation}
commutes.
Therefore, $\Pmathbb_{j \,\cl} (\flat^{*} \Sigma)$ is a canonical internalization of $\Sub_{j \,\cl}(\flat^{*} \Sigma)$.

We can express propositions in different ways.
To do so, we need to invoke the outer presheaf $O$ of D\"{o}ring and Isham \cite{DI08b, DI11,DI12} and a few related notions.
(As for the definition of $O$, see (\ref{eq:outer presheaf 1}) and (\ref{eq:outer presheaf 2}).)
We call the sheafification $\flat^{*} O$ of $O$ the outer sheaf.
Furthermore, we call a set $h \equiv \{\hhat_{V} \in (\flat^{*} O)(V)\}_{V \in \Vbold}$
a hyper-element of $\flat^{*} O$, if
\begin{equation}
\hhat_{V} = \hhat_{\flat(V)}
\quad
\mbox{and}
\quad
\flat^{*} O(V' \hookrightarrow V) (\hhat_{V}) \preceq \hhat_{V'}.
\label{eq:hyper-element of outer sheaf}
\end{equation}
This is a $j$-sheaf counterpart of the notion of hyper-elements (\ref{eq:hyper-element of outer presheaf}) defined by D\"{o}ring and Isham \cite{DI12}.
We write $\Hyp_{j}(\flat^{*}O)$ for the collection of all hyper-elements of $\flat^{*}O$.
Let $\Sub_{j \, \dB} (\flat^{*}O)$ be the collection of all downward closed, Boolean subsheaves of $\flat^{*} O$. That is, for all $P \in \Sub_{j}(\flat^{*}O)$,
$P \in \Sub_{j \, \dB}(\flat^{*}O)$ if and only if, for any $V \in \Vbold$, 
$P(V)$ is a downward closed set of $(\flat^{*} O)(V)$ containing a top element.
(Obviously, such $P(V)$'s are complete Boolean lattices.)
We can regard $\Hyp_{j}(\flat^{*}O)$ and $\Sub_{j \, \dB} (\flat^{*}O)$ as proposition spaces equivalent to $\Sub_{j \, \cl}(\flat^{*} \Sigma)$.
This is because, corresponding to relation (\ref{eq:SubHypSub}) proved by D\"{o}ring and Isham 
\cite{DI12}, the following relation holds:
\begin{equation}
\Sub_{j \; \dB} (\flat^{*}O) \simeq \Hyp_{j} (\flat^{*}O) \simeq \Sub_{j \;{\rm cl}}(\flat^{*}\Sigma).
\label{eq:SubjHypjSubj}
\end{equation}
Here, the bijection at the left hand side of (\ref{eq:SubjHypjSubj}) is realized by a function $c_{j}:\Sub_{j \,\dB} (\flat^{*}O) \xrightarrow{\sim} \Hyp_{j}( \flat^{*}O)$ that is defined by
\begin{equation}
(c_{j}(A))_{V} := \vee  A(V).
\end{equation}
To see the right hand side of  (\ref{eq:SubjHypjSubj}), we use the bijections $\alpha_{V}:
O(V) \to \Cl(\Sigma(V))$ ($V \in \Vbold$) introduced by D\"{o}ring and Isham \cite{DI12}. 
(For the definition, see (\ref{eq:alphaV}).)
These bijections allow us to regard $\{\Cl(\Sigma(V))\}_{V \in \Vbold}$ as a presheaf $\Cl\Sigma$ isomorphic to the outer presheaf $O$,
and $\{ \alpha_{V} \}_{V \in \Vbold}$ as a natural isomorphism $\alpha: O \xrightarrow{\sim} \Cl\Sigma$. 
Furthermore, $\alpha$ induces a natural isomorphism $\flat^{*}\alpha: \flat^{*}O \xrightarrow{\sim} \flat^{*}(\Cl\Sigma) $,
where $\flat^{*}(\Cl\Sigma)(V) = \Cl\Sigma(\flat(V))= \Cl(\Sigma(\flat(V)) ) =  \Cl(\flat^{*} \Sigma (V)) $. 
Therefore, we obtain a bijection $k_{j}: \Hyp_{j} (\flat^{*}O) \xrightarrow{\sim} \Sub_{j \; \cl} (\flat^{*}\Sigma) $ that is given by
\begin{eqnarray}
(k_{j}(h))(V)
& := &
 (\flat^{*}\alpha)_{V}(\hhat_{V}) \nonumber\\
& = &
\{ \sigma \in (\flat^{*}\Sigma)(V) \;|\; \sigma(\hhat_{V}) = 1 \} \nonumber\\
& = &
\{ \sigma \in \Sigma(\flat(V)) \;|\; \sigma(\hhat_{\flat(V)}) = 1 \},
\end{eqnarray}
and hence, a bijection $f_{j}: \Sub_{j \,\dB} (\flat^{*}O) \xrightarrow{\sim} \Sub_{j \; \cl} (\flat^{*}\Sigma)$
defined by
\begin{equation}
(f_{j}(A))(V) := (\flat^{*}\alpha)_{V}(\vee A(V)).
\end{equation}

It is obvious from (\ref{eq:SubjHypjSubj}) that  $k_{j \, \downarrow V}$, $c_{j \, \downarrow V}$,
and $f_{j \, \downarrow V}$, the restrictions of $k_{j}$, $c_{j}$, and $f_{j}$, respectively,
to subalgebras of $V$, give the relation
\begin{equation}
\Sub_{j \; \dB} (\flat^{*}(O_{\downarrow V})) \simeq \Hyp_{j} (\flat^{*}(O_{\downarrow V})) \simeq \Sub_{j \;{\rm cl}}(\flat^{*}(\Sigma_{\downarrow V})).
\end{equation}
Therefore, the proposition space $\Sub_{j \, \cl}(\flat^{*} \Sigma) \simeq \Sub_{j \, \dB}(\flat^{*} O)$ can be internalized also as a subsheaf $\Pmathbb_{j \, \dB} (\flat^{*}O)$ of $\Pmathbb_{j} (\flat^{*}O)$ that is defined by
\begin{equation}
(\Pmathbb_{j \, \dB}(\flat^{*} O))(V) := \Sub_{j \, \dB} (\flat^{*}(O_{\downarrow V})).
\end{equation}
Every proposition $P \in \Sub_{j \, \dB}(\flat^{*} O)$ has  its name $\lceil P \rceil_{j} \in \Gamma(\Pmathbb_{j\, \dB}(\flat^{*} O))$ in $\Sh_{j} \Vboldhat$,
which is given by $(\lceil P \rceil_{j})_{ V} := \flat^{*}(P_{\downarrow V})$.

The daseinization operator $\delta$ introduced by D\"{o}ring and Isham \cite{DI08a,DI08b,DI11,DI12} assigns to each projection operator
$\Phat$ on $\Hcal$ a global element $\delta(\Phat)$ of the outer presheaf $O$.
(For the definition, see (\ref{eq:daseinization}).)
As a counterpart of $\delta$, we introduce a map $\delta_{j}$,
which assigns to each $\Phat$ a global element of $\flat^{*}O$ by
\begin{eqnarray}
\delta_{j}(\Phat)_{V} 
& := &
\bigwedge \{ \ahat \in  (\flat^{*} O)(V) \;|\; \Phat \preceq \ahat \} \nonumber\\
& = &
\bigwedge \{ \ahat \in O(\flat(V))  \;|\; \Phat \preceq \ahat \} \nonumber\\
& = &
\delta(\Phat)_{\flat(V)}.
\end{eqnarray}
To see that really $\delta_{j}(\Phat) \in \Gamma(\flat^{*}O)$,
we note that, for $V' \subseteqVbold V$,
\begin{eqnarray}
(\delta_{j}(\Phat))_{V'} 
& = &
\delta(\Phat)_{\flat(V')} \nonumber \\
& = &
\delta(\delta(\Phat)_{\flat(V)})_{\flat(V')} \nonumber\\
& = &
O(\flat(V') \hookrightarrow \flat(V)) (\delta(\Phat)_{\flat(V)}) \nonumber \\
& = &
(\flat^{*}O)(V' \hookrightarrow V) (\delta_{j}(\Phat)_{V} ).
\label{eq:daseinization and outer sheaf}
\end{eqnarray}
Because of (\ref{eq:SubjHypjSubj}) and the fact that $\Gamma(\flat^{*}O) \subseteq \Hyp_{j}(\flat^{*} O)$,
$\delta_{j}(\Phat)$ can be regarded as a proposition sheaf.
That is, it defines elements of $\Sub_{j \, \dB}(\flat^{*} O)$ and $\Sub_{j \, \cl}(\flat^{*} \Sigma)$ by
\begin{equation}
(\delta_{j}(\Phat) )(V) 
=
\{ \ahat \in (\flat^{*}\Ocal)(V) \;|\; \ahat \preceq (\delta_{j} (\Phat))_{V}  \}
\end{equation}
and
\begin{equation}
(\delta_{j}(\Phat) )(V) =
\{ \sigma \in (\flat^{*} \Sigma)(V) \,|\, 
\sigma ((\delta_{j}(\Phat))_{V})= 1\},
\end{equation} 
respectively.

As previously noted, every proposition is represented by a clopen subsheaf of $\flat^{*} \Sigma$.
We can assign to it a truth-value, a global element of $\Omega_{j}$,
if we are given a collection of truth propositions.
It is internalized as a truth sheaf $\Tmathbb_{j}$,
which is a subsheaf of $\Pmathbb_{j \, \cl}( \flat^{*} \Sigma)$ that satisfies appropriate properties.
We regard a subsheaf $\Tmathbb_{j}$ of $\Pmathbb_{j \,\cl}(\flat^{*}\Sigma)$ as a truth sheaf if and only if $\Tmathbb_{j}(V)$ is a filter for every $V \in \Vbold$.
That is, if $\Tmathbb_{j}(V)$ contains $A  \in \Sub_{j \, \cl} (\flat^{*} (\Sigma_{\downarrow V}))$ as an element,
then it does also any $B \in \Sub_{j \, \cl} (\flat^{*} (\Sigma_{\downarrow V}))$ such that $A \subseteq B$.
Also, if $A$, $B \in \Tmathbb_{j}(V)$,  then $A \cap B \in \Tmathbb_{j}(V)$.

Let $\tau_{j}$ be the characteristic morphism of $\Tmathbb_{j} $
as a subsheaf of $\Pmathbb_{j \, \cl} (\flat^{*}\Sigma)$.
That is, the morphism $\Pmathbb_{j \, \cl} (\flat^{*}\Sigma) \xrightarrow{\tau_{j}} \Omega_{j}$ makes the diagram

\begin{equation}
\xymatrix{
\Tmathbb_{j} \ar [rr] ^{!} \ar @{>->} [dd] && 1 \ar @{>->} [dd] ^{\true_{j}} \\
&& \\
\Pmathbb_{j \, \cl} (\flat^{*}\Sigma) \ar [rr] _{\tau_{j}} && \Omega_{j} \\
}
\label{eq:truth-value pullback_j}
\end{equation}
a pullback.
The morphism $\tau_{j}$ is given by,
for each $S \in \Sub_{j \, \cl}(\flat^{*} (\Sigma_{\downarrow V}))$,
\begin{equation}
(\tau_{j})_{V}(S) = \{ V' \subseteqVbold V \; | \; \flat^{*}(S_{\downarrow V'}) \in \Tmathbb_{j}(V') \}.
\label{eq:characteristic morphism for truth sheaf}
\end{equation}

Given a truth sheaf $\Tmathbb_{j}$,
we can assign to each proposition $P$ a truth value $\nu(P;\Tmathbb_{j}) \in \Gamma \Omega_{j}$ as 
\begin{equation}
\nu(P;\Tmathbb_{j}) = \tau_{j} \circ \lceil P \rceil_{j},
\end{equation}
each $V$-element of which is given by 
\begin{equation}
\nu(P;\Tmathbb_{j}) _{V}
=
\{
V' \subseteqVbold V 
\;|\;
\flat^{*}  (P_{\downarrow V'}) \in \Tmathbb_{j}(V') 
 \}.
\end{equation}

Let $\rho$ be a density matrix and $r \in [0,1]$.
D\"{o}ring and Isham \cite{DI12} defined generalized truth objects $\Tmathbbrhor$,
the definition of which is given by (\ref{eq:truth presheaf rho r}).
Their global elements represent propositions that are only true with probability at least $r$ in the state $\rho$. 
Following D\"{o}ring and Isham, we define $\Tmathbbrhor_{j} \in \Sub_{j\, \dB}(\flat^{*}O)$ by,
for each $V \in \Vbold$,
\begin{equation}
\Tmathbbrhor_{j}(V)
:= \{
A \in \Sub_{j \, \dB}(\flat^{*}(O_{\downarrow V})) \;|\;
\forall V' \subseteqVbold V \;(\tr(\rho (\vee A(V'))) \ge r)
\}.
\label{eq:truth sheaf rho r}
\end{equation}
It is easy to see that every $\Tmathbbrhor_{j}(V)$ is a filter,
and as we will see in proposition \ref{prp:Trhorj}, it is really a $j$-sheaf.

When $r=1$, $\Tmathbb^{\rho ,\, 1}_{j}$ gives propositions that are true in the state $\rho$.
Further, when $\rho=\varphipro$,
$\Tmathbbphi_{j} \equiv \Tmathbb^{\varphipro ,\, 1}_{j}$,
the counterpart of (\ref{eq:truth presheaf for vector state}),
is given by
\begin{eqnarray}
\Tmathbbphi_{j}(V) 
&:=& 
\{ A \in \Sub_{j \, \dB}(\flat^{*}(O_{\downarrow V})) \,|\, \forall V' \subseteqVbold V \; (\delta_{j}(\varphipro)_{V'} \in A(V') ) \}\nonumber\\
& \simeq &
\{ h \in \Hyp_{j} (\flat^{*}(O_{\downarrow \,V}))\;|\; \forall V' \subseteqVbold V \; ( \delta_{j}(\varphipro)_{V'}  \preceq \hhat_{V'} ) \}\nonumber \\
& = &
\{ h \in \Hyp_{j} (\flat^{*}(O_{\downarrow \,V}))\;|\; \forall V' \subseteqVbold V \; (\varphipro \preceq \hhat_{V'} ) \}\nonumber \\
& \simeq &
\{ S \in \Sub_{j \, \cl}(\flat^{*}(\Sigma_{\downarrow V})) \,|\, \forall V' \subseteqVbold V \;
(\varphipro \preceq (\flat^{*}\alpha)_{V'}^{-1} (S(V')) )\} . \nonumber\\
\label{eq:truth sheaf for vector state}
\end{eqnarray}
\begin{prp}
\label{prp:Trhorj}
For every state $\rho$ and every coefficient $r \in [0,1]$,  
$\Tmathbbrhor_{j}$ is a $j$-sheaf.
\end{prp}
Proof.
First, we show that $\Tmathbbrhor_{j}$ is a presheaf.
To do so,
for each $V \in \Vbold$, let $V' \subseteqVbold V$ and $V'' \subseteqVbold V'$.
Then, since we have $\flat(V'') \subseteqVbold V'$,
it follows that for any $A \in \Sub_{j \, \dB}(\flat^{*}(O_{\downarrow V}))$,
\begin{equation}
\flat^{*}(A_{\downarrow V'})(V'') = A_{\downarrow V'}(\flat(V'')) = A(\flat(V'')),
\end{equation}
hence,
\begin{equation}
\tr(\rho(\vee \flat^{*}(A_{\downarrow V'})(V''))) = \tr(\rho (\vee A(\flat(V'')))).
\end{equation}
This means that $A \in \Tmathbbrhor_{j}(V)$ implies $\flat^{*}(A_{\downarrow V'}) \in \Tmathbbrhor_{j}(V')$,
that is, $\Tmathbbrhor_{j}$ is a presheaf.

Next, let $A \in (\flat^{*}\Tmathbbrhor_{j})(V)=\Tmathbbrhor_{j}(\flat(V))$;
that is, suppose that for every $V' \subseteq \flat(V)$, $\tr(\rho (\vee A(V'))) \ge r$.
Then, for every $V' \subseteqVbold V$, since $\flat(V') \subseteqVbold \flat(V)$,
we have 
\begin{equation}
\tr(\rho (\vee A(V'))) = \tr(\rho (\vee A(\flat(V')))) \ge r,
\end{equation}
which means $A \in \Tmathbbrhor_{j}(V)$. 
Thus, $ (\flat^{*}\Tmathbbrhor_{j})(V) \subseteq \Tmathbbrhor_{j}(V)$ results.

Finally, $\Tmathbbrhor \xrightarrow{\zeta_{\Tmathbbrhor_{j}}} \flat^{*}\Tmathbbrhor$ turns out to be a natural isomorphism, 
because for every $A \in \Tmathbbrhor_{j}(V)$, 
$(\zeta_{\Tmathbbrhor_{j}})_{V}(A) = \flat^{*}(A_{\downarrow \flat(V)}) = \flat^{*}(A_{\downarrow V}) = A$. \qed

Let $\Pmathbb_{j \, \dB}(\flat^{*}O) \xrightarrow{\taurhor_{j}} \Omega_{j}$ be the characteristic morphism of $\Tmathbbrhor_{j} $.
From (\ref{eq:characteristic morphism for truth sheaf}), we have, for each $A \in \Sub_{j \, \dB}(\flat^{*}(O_{\downarrow V})) = (\Pmathbb_{j \, \dB}(\flat^{*}O))(V)$,
\begin{eqnarray}
(\taurhor_{j})_{V}(A) 
& = &
\{  V' \subseteqVbold V \;|\; 
\forall V'' \subseteqVbold V' \; (\tr(\rho (\vee \flat^{*}(A_{\downarrow V'})(V''))) \ge r)
\} \nonumber\\
& = &
\{  V' \subseteqVbold V \;|\; 
\forall V'' \subseteqVbold V' \; (\tr(\rho (\vee A(V''))) \ge r)
\}.
\end{eqnarray}
Therefore, the truth-value of a physical proposition $\delta_{j}(\Phat)$ corresponding to a projection operator $\Phat$ under the truth sheaf $\Tmathbbrhor$ is given by, for each $V \in \Vbold$,
\begin{eqnarray}
\nu_{j}(\delta_{j}(\Phat);\Tmathbbrhor_{j})_{V}
& = &
\{ V' \subseteqVbold V \;|\;
\forall V'' \subseteqVbold V' \; (\tr (\rho (\delta_{j}(\Phat)_{V''}) )\ge r)
\} \nonumber\\
& = &
\{ V' \subseteqVbold V \;|\;
\tr (\rho (\delta_{j}(\Phat)_{V'}) )\ge r
\}.
\end{eqnarray}
In particular, for   $\Tmathbbrhor_{j} = \Tmathbbphi_{j}$, we have
\begin{eqnarray}
\nu_{j}(\delta_{j}(\Phat);\Tmathbbphi_{j})_{V}
& = &
\{ V' \subseteqVbold V \;|\;
\langle \varphi | (\delta_{j}(\Phat)_{V'}) \varphivec = 1
\} \nonumber\\
& = &
\{ V' \subseteqVbold V \;|\; \varphipro \preceq \delta_{j}(\Phat)_{V'} \}\nonumber\\
& = &
\{ V' \subseteqVbold V \;|\; \delta_{j}(\varphipro) \in \delta_{j}(\Phat)(V') \}.
\label{eq:nu_j 2}
\end{eqnarray}


\section{Translation Rules of Propositions, Truth Objects, and Truth-Values}
\label{sec:translation rule}

In Section \ref{sec:truth-value valuation},
we gave the truth-value function $\nu_{j}$ that assigns a truth-value to each proposition sheaf $P_{j}$ under a given truth sheaf $\Tmathbb_{j}$.
In this and the next sections, we clarify the structural relationship between the present sheaf-based theory and the presheaf-based one.
What we show in this section is that, for each $P_{j}$ and $\Tmathbb_{j}$,
there are corresponding proposition presheaves $P$ and truth presheaves $\Tmathbb$
that can be regarded as `translations',
and that there exists a specific relation between global elements of $\Omega_{j}$ and $\Omega$,
which is satisfied by $\nu_{j}(P_{j};\Tmathbb_{j})$ and $\nu(P;\Tmathbb)$ for all such propositions $P_{j}$
and $P$ and truth objects $\Tmathbb_{j}$ and $\Tmathbb$.
Precisely, we show that they satisfy the following relation:
\begin{equation}
\nu_{j}(P_{j};\Tmathbb_{j}) =r \circ \nu(P;\Tmathbb),
\label{eq:relation}
\end{equation}
where the morphism $r$ is defined by the epi-mono factorization of $j$,
\begin{equation}
\xymatrix{
\Omega \ar [rr] ^{j} \ar @{->>} [rdd] _{r} && \Omega \, ,\\
&& \\
& \Omega_{j}  \ar @{>->} [ruu]  \\
}
\label{eq:epi-mono}
\end{equation}
that is, $r$ is defined by $r_{V}(\omega) \equiv j_{V}(\omega) \in \Omega_{j}(V)$.
In the following, we give concrete translation relationships for proposition objects $P$ and $P_{j}$ and for truth objects $\Tmathbb$ and $\Tmathbb_{j}$.

First, we give a definition of translation of propositions.
Note that each proposition presheaf $P \in \Sub_{\dB}(O)$ is sheafificated to a proposition sheaf $\flat^{*} P \in \Sub_{j \, \dB}(\flat^{*} O)$.
Therefore, it is quite natural to regard $P$ and $P_{j}$ as each other's translation if they satisfy 
\begin{equation}
\flat^{*} P = P_{j}.
\label{eq:translation condition of proposition}
\end{equation}
The following proposition, which is clear from the definition of $\delta_{j}(\Phat)$, would suggest (\ref{eq:translation condition of proposition}) as a sound definition of translation.
\begin{prp}
For every projection operator $\Phat$, $\delta(\Phat)$ and $\delta_{j}(\Phat)$ are each other's translations.
\end{prp}

Next, we define translation of truth objects.
First, we note that, for each truth sheaf $\Tmathbb_{j} \in \Sub_{j} (\Pmathbb_{j \, \dB}(\flat^{*} O))$, the morphism $\flat^{*}(\Pmathbb_{\dB} O) \xrightarrow{\varrho_{O}} \Pmathbb_{j \, \dB}(\flat^{*} O)$ induces a subsheaf of $\flat^{*}(\Pmathbb_{\dB} O) $, which we denote by $\varrho_{O}^{-1}(\Tmathbb_{j})$,
as the pullback of $\Tmathbb_{j}$ along the morphism $\varrho_{O}$; that is,
\begin{eqnarray}
\varrho_{O}^{-1}(\Tmathbb_{j}) (V)
& := & 
(\varrho_{O})_{V}^{-1}(\Tmathbb_{j}(V) ) \nonumber\\
& = &
\{ A \in (\flat^{*}(\Pmathbb_{\dB} O) )(V) 
\;|\;
\flat^{*} A \in \Tmathbb_{j}(V)  \} \nonumber\\
& = & 
\{ A \in  \Sub_{\dB}(O_{\downarrow \flat(V)})
\;|\;
\flat^{*} A \in \Tmathbb_{j}(V)  \}.
\label{eq:intermediate translation of truth sheaf}
\end{eqnarray}
On the other hand, each truth presheaf $\Tmathbb \in \Sub(\Pmathbb_{\dB} O)$,
for which we propose that $\Tmathbb(V)$ is a filter for every $V \in \Vbold$,
has its sheafification $\flat^{*} \Tmathbb \in  \Sub_{j} (\flat^{*} (\Pmathbb_{\dB} O))$.
We say that $\Tmathbb$ and $\Tmathbb_{j}$ are each other's translation,
if they satisfy
\begin{equation}
\flat^{*} \Tmathbb = \varrho_{O}^{-1} (\Tmathbb_{j}) .
\label{eq:translation condition of truth sheaf}
\end{equation}

To show soundness of the definition (\ref{eq:translation condition of truth sheaf}) of translation, 
we give the following proposition.
\begin{prp}
For every density matrix $\rho$ and $r \in [0,1]$,
the corresponding truth presheaf $\Tmathbbrhor$ and the truth sheaf $\Tmathbbrhor_{j}$ are each other's translation.
\end{prp}  
Proof.
Let $A \in \Sub_{\dB}(O_{\downarrow \flat(V)})$ and $h \in \Hyp(O_{\downarrow \flat(V)})$ be its corresponding hyper-element.
Then $A \in (\flat^{*}\Tmathbbrhor)(V)$ if and only if
\begin{equation}
\forall V' \subseteqVbold \flat(V) \quad
\tr(\rho \hhat_{V'}) \ge r
\label{eq:sheafification of truth presheaf},
\end{equation}
whereas $A \in (\varrho_{O}^{-1}(\Tmathbbrhor_{j}))(V)$ if and only if
\begin{equation}
\forall V' \subseteqVbold V \quad
\tr(\rho \hhat_{\flat(V')}) \ge r.
\label{eq:intermediate translation of truth sheaf for vector state}
\end{equation}
What we have to prove is that 
(\ref{eq:sheafification of truth presheaf}) and (\ref{eq:intermediate translation of truth sheaf for vector state}) are equivalent.

Suppose that (\ref{eq:sheafification of truth presheaf}) holds.
Then, since for $V' \subseteqVbold V$, we have $\flat(V') \subseteqVbold \flat(V)$,
(\ref{eq:intermediate translation of truth sheaf for vector state}) follows.

Conversely, suppose that  (\ref{eq:intermediate translation of truth sheaf for vector state}).
Then, in particular, 
\begin{equation}
\tr(\rho \hhat_{\flat(V)}) \ge r.
\end{equation}
On the other hand, for every $V' \subseteqVbold \flat(V)$,
\begin{equation}
\hhat_{\flat(V)} \preceq \delta(\hhat_{\flat(V)})_{V'} \preceq \hhat_{V'}.
\end{equation}
Thus, we have
\begin{equation}
r \le \tr(\rho \hhat_{\flat(V)})  \le \tr(\rho \hhat_{V'}) ,
\end{equation}
which implies (\ref{eq:sheafification of truth presheaf}).
\qed

Now, let $P$ and $\Tmathbb$ be arbitrary translations of $P_{j}$ and $\Tmathbb_{j}$, respectively.
In the following, we prove that they really satisfy (\ref{eq:relation}). 

First, note that the names $\lceil P \rceil$ and $\lceil P_{j} \rceil_{j}$ make the diagram
\begin{equation}
\xymatrix{
  &&  \Pmathbb_{\dB} O \ar [d] ^{\zeta_{\Pmathbb_{\dB} O}}\\
1  \ar @{>->} [rru] ^{\lceil P \rceil} \ar @{>->} [rrd] _{\lceil P_{j} \rceil_{j}} &&  \flat^{*}(\Pmathbb_{\dB} O)
\ar [d] ^{\varrho_{O}} \\
  && \Pmathbb_{j \, \dB} (\flat^{*} O) \\
}
\label{eq:triangle of names}
\end{equation}
commute.
Here, the definition of 
$\flat^{*}(\Pmathbb_{\dB} O) \xrightarrow{\varrho_{O}} \Pmathbb_{j \, \dB}(\flat^{*}O)$, the restriction of $\flat^{*}(\Pmathbb  O) \xrightarrow{\varrho_{O}} \Pmathbb_{j}(\flat^{*}O)$, is given in appendix \ref{sec:misc}.
The commutativity of (\ref{eq:triangle of names}) is easily shown as
\begin{eqnarray}
(\lceil P_{j} \rceil_{j} )_{V}
& = &
\flat^{*}((\flat^{*} P)_{\downarrow V}) \nonumber\\
& = &
\flat^{*}(P_{\downarrow V}) \nonumber\\
& = &
\flat^{*}((P_{\downarrow V})_{\downarrow V}) \nonumber\\
& = &
\flat^{*}((P_{\downarrow V})_{\downarrow \flat(V)}) \nonumber\\
& = &
(\varrho_{O})_{V}((\zeta_{\Pmathbb_{\dB} O})_{V}(\lceil P \rceil_{V})),
\end{eqnarray}
where we used (\ref{eq:flat1}) and (\ref{eq:flat3}).

\begin{prp}
\label{prp:translation condition diagram of truth sheaf}
Let $\Pmathbb_{\dB} O \xrightarrow{\tau} \Omega$ and 
$\Pmathbb_{j \, \dB} (\flat^{*}O) \xrightarrow{\tau_{j}} \Omega_{j}$ be the characteristic morphisms of $\Tmathbb$ and $\Tmathbb_{j}$, respectively.
Then, the diagram
\begin{equation}
\xymatrix{
\flat^{*} (\Pmathbb_{\dB} O) \ar [rr] ^{\flat^{*} \tau} \ar [dd] _{\varrho_{O}} && \flat^{*} \Omega \ar  [d] ^{\flat^{*} r} \\
&& \flat^{*} \Omega_{j} \\
\Pmathbb_{j \, \dB}(\flat^{*} O) \ar [rr] _{\tau_{j}}      && \Omega_{j}  \ar [u] ^{\vertsim} _{\zeta_{\Omega_{j}}}  \\
} 
\label{eq:translation condition diagram of truth sheaf}
\end{equation}
commutes if and only if 
equation (\ref{eq:translation condition of truth sheaf}) is satisfied.
\end{prp}
Proof.
First, we note that, for each $A \in \flat^{*}(\Pmathbb_{\dB} O)(V) = \Sub_{\dB}(O_{\downarrow \flat(V)})$,
\begin{eqnarray}
(\flat^{*}r \circ \flat^{*}\tau)_{V}(A)
& = &
\{
V' \subseteqVbold \flat(V) \;|\;
\flat(V') \in \tau_{\flat(V)}(A)
\} \nonumber \\
&=&
\{
V' \subseteqVbold \flat(V) \;|\;
A_{\downarrow \flat(V')} \in \Tmathbb(\flat(V'))
\},
\end{eqnarray}
and,
\begin{equation}
(\zeta_{\Omega_{j}} \circ \tau_{j} \circ \varrho_{O})_{V}(A) 
=
\{
V' \subseteqVbold \flat(V) \;|\;
\flat^{*}(A_{\downarrow V'}) \in \Tmathbb_{j}(V')
\}.
\end{equation}
Suppose that the diagram (\ref{eq:translation condition diagram of truth sheaf}) commutes.
Then, for each $V' \subseteqVbold \flat(V)$,
$A_{\downarrow \flat(V')} \in \Tmathbb(\flat(V'))$
if and only if 
$\flat^{*}(A_{\downarrow V'}) \in \Tmathbb_{j}(V')$.
In particular, putting $V' = \flat(V)$,
we obtain equation (\ref{eq:translation condition of truth sheaf}).
Conversely, suppose that (\ref{eq:translation condition of truth sheaf}) holds.
Then, we have, for each $V \in \Vbold$ and $V' \subseteqVbold \flat(V)$,
$\flat^{*} \Tmathbb (V') = \varrho_{O}^{-1}(\Tmathbb_{j})(V')$;
that is, for all $A' \in \Sub (O_{\downarrow \flat(V')})$, 
$A' \in \Tmathbb(\flat(V'))$ if and only if $\flat^{*} A' \in \Tmathbb_{j}(V')$.
In particular, for any $A \in \Sub_{\dB}(O_{\downarrow \flat(V)})$,
we obtain the condition for the diagram (\ref{eq:translation condition diagram of truth sheaf}) to commute, by taking $A' = A_{\downarrow \flat(V')}$.
\qed

To show the relation (\ref{eq:relation}),
let $\Tmathbb$ and $P$ be transforlations of $\Tmathbb_{j}$ and $P_{j}$, respectively.

Fitting together (\ref{eq:triangle of names}), (\ref{eq:translation condition diagram of truth sheaf}), and naturality of $\zeta$,
we have a commutative diagram
\begin{equation}
\xymatrix{
&&\Pmathbb_{\dB} O \ar [rrr] ^{\tau}  \ar  [dd] _{\zeta_{\Pmathbb_{\dB} O}}&& & \Omega \ar  [ddll] ^{\zeta_{\Omega}} \ar [dddd] ^{r} \\
&& &&& \\
1\ar @{>->}[rruu] ^{\lceil P \rceil} \ar @{>->}[rrdd] _{\lceil P_{j} \rceil _{j}} && \flat^{*}(\Pmathbb_{\dB} O) \ar [r] ^(0.6){\flat^{*} \tau} \ar [dd] _{\varrho_{O}}
 &  \flat^{*}\Omega \ar [dr] ^{\flat^{*} r} &&& \\
&&&& \flat^{*} \Omega_{j} \ar [dr] ^{\zeta^{-1}_{\Omega_{j}}} _{\fortyfivesim}  \\
&&\Pmathbb_{j \, \dB} (\flat^{*}O) \ar [rrr] _{\tau_{j}}   && & \Omega_{j}  .\\
}
\label{eq:tbox}
\end{equation}
The outer pentagon of this diagram is just the relation (\ref{eq:relation}).

We have proved that 
for all proposition objects $P$ and $P_{j}$ satisfying (\ref{eq:translation condition of proposition}) and truth objects $\Tmathbb$ and $\Tmathbb_{j}$ satisying (\ref{eq:translation condition of truth sheaf}), 
the truth-values $\nu(P,\Tmathbb)$ and $\nu_{j}(P_{j},\Tmathbb_{j})$ are related via (\ref{eq:relation}).
This implies that $P$ and $\Tmathbb$ represent virtually the same proposition as $P_{j}$ and the same truth object as $\Tmathbb_{j}$, respectively, from our sheaf-based viewpoint.
In this sense, it is reasonable to call them each other's translation.
Also, we call the same relation between global elements of $\Omega_{j}$ and $\Omega$ as (\ref{eq:relation}), that is,
\begin{equation}
\nu_{j} = r \circ \nu,
\label{eq:translation relation of truth-value}
\end{equation}
the translation rule of global elements,
and say that $\nu_{j} \in \Gamma\Omega_{j}$ and $\nu \in \Gamma \Omega$ are each other's translation if they satisfy (\ref{eq:translation relation of truth-value}).

%
%


\section{Coarse-Graining Properties of Translation}
\label{sec:coarse-graining}

For a proposition $P_{j}$ and a truth sheaf and $\Tmathbb_{j}$,
their translation presheaves $P$ and $\Tmathbb$ satisfying (\ref{eq:translation condition of proposition}) and (\ref{eq:translation condition of truth sheaf}) are not determined uniquely.
For such $P$'s and $\Tmathbb$'s, furthermore, the truth-values $\nu(P,\Tmathbb)$ take various values.
If we consider their sheaf translations, the various truth-values are transformed to one and the same value $r \circ \nu(P, \Tmathbb)$.
In other words, a lot of different propositions, truth objects, and truth-values are not distinguished from the sheaf-based viewpoint.
We call this aspect coarse-graining made by translation,
the properties of which we observe in the following.

First, let us see coarse-graining of the space $\Gamma \Omega$ of truth-values.
The translation rule (\ref{eq:translation relation of truth-value}) is equivalent to the condition that
for all $V \in \Vbold$ and $V' \subseteqVbold V$,
\begin{equation}
V' \in (\nu_{j})_{V} \iff \flat(V') \in \nu_{V}.
\label{eq:translation condition of truth-value}
\end{equation}
Let $\gammabold(\nu_{j})$ be the set of all translations $\nu \in \Gamma \Omega$ of $\nu_{j}$.
Note that $\gammabold(\nu_{j})$ has an order relation $\le$ inherited from $\Gamma \Omega$.
Namely, $\nu_{1} \le \nu_{2}$ if and only if $(\nu_{1})_{V} \subseteq (\nu_{2})_{V}$ for all $V \in \Vbold$. 
Furthermore, $\gammabold(\nu_{j})$ is closed with respect to binary operations on $\Gamma \Omega$,
the join $\vee$ and the meet $\wedge$, 
each of which is defined by $(\nu_{1} \vee \nu_{2})_{V} := (\nu_{1})_{V} \cup (\nu_{2})_{V} $ and $(\nu_{1} \wedge \nu_{2})_{V} := (\nu_{1})_{V} \cap (\nu_{2})_{V} $, respectively.

Let us define $\gammaboldmax(\nu_{j}) \in \Gamma \Omega$ by
\begin{equation}
\gammaboldmax(\nu_{j}) :=\begin{xy} { (0,0)*{1_{}\;} \ar @{>->} (10,0)*{\;\Omega_{j}}^(0.4){\nu_{j}} \ar @{>->} (15,0);(23,0) *{\; \Omega}} \end{xy}.
\end{equation}
This is the maximum translation of $\nu_{j}$.
In fact, it is clear from the definition (\ref{eq:subobject classifier of quantization topos}) that (\ref{eq:translation condition of truth-value}) is satisfied if we put $\nu =\gammaboldmax(\nu_{j})$.
Furthermore, if $\nu \in \gammabold(\nu_{j})$,
then, since $\flat(V') \in \nu_{V}$ for every $V' \in \nu_{V}$, we have $V' \in (\nu_{j})_{V}=\gammaboldmax(\nu_{j})_{V}$ from (\ref{eq:translation condition of truth-value}).
Thus, $\nu \le \gammaboldmax(\nu_{j})$ follows.

Let us define $\gammaboldmin(\nu_{j})$ by
%
%
\begin{equation}
\gammaboldmin(\nu_{j})(V)
:= \{
V' \subseteqVbold V \;|\;
 (\nu_{j})_{V} \cap \Ucalflat(V') \ne \emptyset \},
\end{equation}
where, for each $V \in \Vbold$, $\Ucalflat(V)$ is defined by
\begin{equation}
\Ucalflat(V) := \{W \in \Vbold \;|\; V \subseteqVbold \flat(W) \}.
\end{equation}
We can straightforwardly verify that $\gammaboldmin(\nu) \in \gammabold(\nu_{j})$.
Moreover, $\gammaboldmin(\nu_{j})$ is the least translation of $\nu_{j}$.
To see this, let $\nu \in \gammabold(\nu_{j})$ and $V' \in \gammaboldmin(\nu_{j})_{V}$.
Then, we have $V' \subseteqVbold \flat(V'')$ for some $V'' \in (\nu_{j})_{V}$.
Furthermore, since for such $V''$, $\flat(V'') \in \nu_{V}$ because of (\ref{eq:translation condition of truth-value}).
Thus, we have $V' \in \nu_{V}$ since $V' \subseteqVbold \flat(V'')$.
Conversely, it is easy to show that every $\nu \in \Gamma \Omega$ lying between $\gammaboldmin(\nu_{j})$ and $\gammaboldmax(\nu_{j})$ satisfies (\ref{eq:translation condition of truth-value}).
On the other hand, every $\nu \in \Gamma \Omega$ is a translation of $r \circ \nu \in \Gamma \Omega_{j}$.
We thus obtain the following result:
\begin{thm}
\label{thm:coarse-graining of truth-value}
The truth-value space $\Gamma \Omega$ can be expressed as a disjoint union of the lattices $\gammabold(\nu_{j})$ ($\nu_{j} \in \Gamma_{j} \Omega_{j}$),
each of which is given by
\begin{equation}
\gammabold(\nu_{j}) = \{ \nu \in \Gamma \Omega\;|\; \gammaboldmin(\nu_{j}) \le \nu \le \gammaboldmax(\nu_{j}) \}.
\end{equation}
\end{thm}

Next, let us turn to the definition (\ref{eq:translation condition of proposition}) of translation of propositions.
Let $\imathbold(P_{j})$ be the set of all translation presheaves of $P_{j}$.
It is clear that $\imathbold(P_{j})$ is an ordered set with respect to the inclusion relation defined on $\Sub_{\cl} \Sigma \simeq \Sub_{\dB} O$.
That is, $P_{1} \subseteq P_{2}$ if and only if $P_{1}(V) \subseteq P_{2}(V)$ for all $V \in \Vbold$.
Also, since (\ref{eq:translation condition of proposition}) is equivalent to 
$P({\flat(V)}) = P_{j}(V)$ for all $V \in \Vbold$,
$\imathbold(P_{j})$ is closed for $\vee$ and $\wedge$ defined on $\Sub_{\cl} \Sigma$,
where $P_{1} \wedge P_{2}$ and $P_{1} \vee P_{2}$ are defined by 
$(P_{1} \wedge P_{2})(V) := P_{1}(V) \cap P_{2}(V)$ and $(P_{1} \vee P_{2})(V) := P_{1}(V) \cup P_{2}(V)$, respectively.

Among the translations $P \in \imathbold(P_{j})$, 
there exists a canonical one $\imathboldmax(P_{j})$.
To give the definition, we note the following fact.
\begin{prp}
If $h \equiv \{ \hhat_{V} \in (\flat^{*} O)(V) \}_{V \in \Vbold}$ is a hyper-element of $\flat^{*}O$,
it is also a hyper-element of $O$.
\end{prp}
Proof.
Let $h$ be a hyper-element of  $\flat^{*}O$.
Then, since 
$\flat(V) \subseteqVbold V$ for every $V \in \Vbold$, 
we have $\hhat_{V} \in (\flat^{*}O)(V) = O(\flat(V)) \subseteq O(V)$,
whereas we have $\flat(V') \subseteqVbold V'$ for every $V' \subseteqVbold V$.
Thus, from (\ref{eq:hyper-element of outer sheaf}) and (\ref{eq:daseinization and outer sheaf}),  it follows
\begin{eqnarray}
\delta(\hhat_{V})_{V'}  \preceq 
\delta(\delta(\hhat_{V})_{V'} )_{\flat(V')}=
\delta(\hhat_{V})_{\flat(V')} = 
\delta_{j}(\hhat_{V})_{V'} 
 \preceq 
\hhat_{V'},
\end{eqnarray}
which means that $h$ is a hyper-element of $O$.
\qed

Every proposition sheaf $P_{j} \in \Sub_{j \, \dB}(\flat^{*} O)$ has its hyper-element $\{\vee P_{j}(V) \}_{V \in \Vbold}$ of $\flat^{*}O$.
We define $\imathboldmax(P_{j})$ as the proposition presheaf given by $\{\vee P_{j}(V) \}_{V \in \Vbold}$ as a hyper-element of $O$:
\begin{eqnarray}
\imathboldmax (P_{j}) (V)
& := &  \{ \ahat \in O(V) \;|\; \ahat \preceq  \vee P_{j}(V) \}  \nonumber\\
& = & \{ \ahat \in O(V) \;|\; \delta(\ahat)_{\flat(V)} \in P_{j}(V) \}  \nonumber\\
& = &  ((\zeta_{O})_{V})^{-1}(P_{j}(V)).
\label{eq:imathboldPj}
\end{eqnarray}
Clearly, $\imathboldmax (P_{j})$ satisfies  $\flat^{*}(\imathboldmax (P_{j}))= P$, that is, it is really a translation of $P_{j}$.
\begin{prp}
For every proposition sheaf $P_{j} \in \Sub_{j\, \dB}(\flat^{*}O)$, $\imathboldmax (P_{j})$ is the largest translation of $P_{j}$.
\end{prp}
Proof.
The third line of (\ref{eq:imathboldPj}) means that $\imathboldmax (P_{j})$ is a pullback of $\xymatrix{ \ar @{>->}(0,0)*{P_{j}\;\,};(10,0) *{\;\flat^{*} O_{}}}$ along the morphism $O \xrightarrow{\zeta_{O}}  \flat^{*}O$.
That is, it makes the tropezoid in the diagram
\begin{equation}
\xymatrix{
 P  \ar @{>->} [rr]  \ar [dd] _{\zeta_{P}} &&  O \ar [dd] ^{\zeta_{O}}\\
 &  \imathboldmax(P_{i}) \ar @{>->} [ru] \ar [d] & \\
\flat^{*}P \ar @{=} [r] & P_{j} \ar @{>->} [r] & \flat^{*} O \\
}
\end{equation}
a pullback.
On the other hand, if $\flat^{*} P = P_{j}$, the outer square commutes because of naturality of $\zeta$.
We thus obtain an inclusion 
$\xymatrix{ \ar @{>->}(0,0)*{P\;};(13,0) *{\;\imathboldmax(P_{j})}}$.
\qed

For instance, for every projection operator $\Phat$, 
we have $ \delta(\Phat)_{V} \preceq \delta(\Phat)_{\flat(V)} = \delta_{j}(\Phat)_{V}$,
whereas $\delta_{j}(\Phat)$ defines $\imathboldmax(\delta_{j}(\Phat))$ as a hyper element of $O$.
Therefore, the proposition presheaf $\delta(\Phat)$, 
which is a translation of $\delta_{j}(\Phat)$ as previously mentioned, 
is included by $\imathboldmax(\delta_{j}(\Phat))$.

We define $\imathboldmin(P_{j})$ by, for each $V \in \Vbold$,
\begin{equation}
\imathboldmin(P_{j})(V) :=
\begin{cases}
\{ \ahat \in O(V) \;|\; \ahat \preceq \bigvee \{ \delta(\vee P_{j}(W))_{V} \}_{W \in \Ucalflat(V)} \} & \mbox{if }\, \Ucalflat(V) \ne \emptyset, \\
\{ \Ohat \} & \mbox{if }\, \Ucalflat(V) = \emptyset.
\end{cases}
\end{equation}
\begin{prp}
For every proposition sheaf $P_{j}$,
$\imathboldmin(P_{j})$ is the smallest translation of $P_{j}$.
\end{prp}
Proof.
Let $k \in \Hyp_{j}(\flat^{*}O)$ be the hyper-element corresponding to $P_{j}$.
To show that $\imathboldmin(P_{j})$ is a presheaf,
we define $h := \{ \hhat_{V} \in O(V)\}_{V \in \Vbold}$ by
\begin{equation}
\hhat_{V}
:=
\begin{cases}
\bigvee\{ \delta(\khat_{W})_{V} \}_{W \in \Ucalflat(V)} & \mbox{if }\,  \Ucalflat(V) \ne \emptyset, \\
 \Ohat  &  \mbox{if }\, \Ucalflat(V) = \emptyset.
\end{cases}
\end{equation}
Since for each $V$, $\hhat_{V}$ is the top element of $\imathboldmin(P_{j})(V)$,
$\imathboldmin(P_{j})$ is a presheaf if and only if $h$ is a hyper-element of $O$.
Let us show this, first.

Suppose that $\Ucalflat(V) \ne \emptyset$.
Since $\Ucalflat(V) \subseteq \Ucalflat(V')$ for every $V' \subseteqVbold V$,
we have
\begin{equation}
\hhat_{V}= \bigvee\{ \delta(\khat_{W})_{V} \}_{W \in \Ucalflat(V)}
\preceq \bigvee\{ \delta(\khat_{W'})_{V} \}_{W' \in \Ucalflat(V')}.
\label{eq:presheaf proof 1}
\end{equation}
On the other hand, since $O(V') \subseteq O(V)$,
we have, for every $W' \in \Ucalflat(V')$,
\begin{equation}
\delta(\khat_{W'})_{V} \preceq \delta(\khat_{W'})_{V'},
\end{equation}
hence,
\begin{equation}
\bigvee\{\delta(\khat_{W'})_{V}\}_{W' \in \Ucalflat(V')} \preceq 
\bigvee\{\delta(\khat_{W'})_{V'}\}_{W' \in \Ucalflat(V')} = \hhat_{V'}.
\label{eq:presheaf proof 2}
\end{equation}
From (\ref{eq:presheaf proof 1}) and (\ref{eq:presheaf proof 2}),
we have
\begin{equation}
\delta(\hhat_{V})_{V'} \preceq \delta(\hhat_{V'})_{V'} = \hhat_{V'}.
\end{equation}
If $\Ucalflat(V) = \emptyset$, then $\delta (\hhat^{V})_{V'} = \Ohat \preceq \hhat_{V'} $.
Thus, $h$ is a hyper-element of $O$, hence really, $\imathboldmin(P_{j})$ a presheaf.

In order to show that $\imathboldmin(P_{j}) \in \imathbold(P_{j})$,
it suffices to show that $\hhat_{\flat(V)} = \khat_{V}$ for every $V \in \Vbold$.
Since $V \in \Ucalflat(\flat(V))$,
we have
\begin{eqnarray}
\hhat_{\flat(V)}
& = &
\bigvee\{ \delta(\khat_{W})_{\flat(V)}\}_{W \in \Ucalflat(\flat(V))} \nonumber\\
& = &
(\delta(\khat_{V})_{\flat(V)}) \vee (\bigvee\{ \delta(\khat_{W})_{\flat(V)}\}_{W \in \Ucalflat(\flat(V)) \setminus \{V\}}).
\end{eqnarray}
On the other hand, we have $\delta(\khat_{V})_{\flat(V)} = \delta_{j}(\khat_{\flat(V)})_{\flat(V)} =\khat_{\flat(V)}= \khat_{V}$,
and $\delta(\khat_{W})_{\flat(V)} \preceq \khat_{\flat(V)} = \khat_{V}$ for all $W \in \Ucalflat(\flat(V)) \setminus \{V\}$.
Thus, $\hhat_{\flat(V)} = \khat_{V}$ results.

Finally, to show $\imathboldmin(P_{j})$ to be the smallest translation of $P_{j}$,
let $l \in \Hyp(O)$ be the hyper-element corresponding to a translation $P \in \imathbold(P_{j})$.
What we have to show is $\hhat_{V} \preceq \lhat_{V}$ for all $V \in \Vbold$.
It suffices to treat the case where $\Ucalflat(V) \ne \emptyset$.
Since we have
\begin{equation}
\delta(\khat_{W})_{V} = \delta(\lhat_{\flat(W)})_{V} \preceq \lhat_{V}
\end{equation}
for all $W \in \Ucalflat(V)$,
it follows that
\begin{equation}
\hhat_{V}
=
\bigvee \{ \delta(\khat_{W})_{V} \}_{W \in \Ucalflat(V)} \preceq 
\lhat_{V}.
\end{equation}
\qed

It is obvious that every proposition presheaf $P \in \Sub_{\dB}(O)$ is a translation of $P_{j}$ if and only if $\imathboldmin(P_{j}) \subseteq P \subseteq \imathboldmax(P_{j})$.
On the other hand, every proposition presheaf $P$ is a translation of the proposition sheaf $\flat^{*}P$.
Thus, we obtain the following result.
\begin{thm}
\label{thm:coarse-graining of proposition}
The proposition space $\Sub_{\dB}(O)$ can be expressed as a disjoint union of the lattices $\imathbold(P_{j})$ ($P_{j} \in \Sub_{j \, \dB}(\flat^{*}O)$), 
each of which is given by
\begin{equation}
\imathbold(P_{j}) = \{ P \in \Sub_{\cl} (\Sigma )\;|\; \imathboldmin(P_{j}) \subseteq P \subseteq \imathboldmax(P_{j})\}.
\end{equation}
\end{thm}

Finally, we observe coarse-graining of truth presheaves.
Let $\Tpre$ be the set of all truth presheaves;
that is,  $\Tmathbb \in \Tpre$ means that $\Tmathbb \in \Sub(\Pmathbb_{\cl} \Sigma)$ and $\Tmathbb(V)$ is a filter for every $V \in \Vbold$.
We first note that we can define $\vee$ and $\wedge$ on $\Tpre$.
In fact, we define $\Tmathbb_{1} \vee \Tmathbb_{2} := \Tmathbb_{1} \cap \Tmathbb_{2}$,
whereas for $\Tmathbb_{1} \wedge \Tmathbb_{2}$, we let
$(\Tmathbb_{1} \vee \Tmathbb_{2})(V)$ the smallest filter $\Fcal(\Tmathbb_{1}(V) \cup \Tmathbb_{2}(V))$ including $\Tmathbb_{1}(V) \cup \Tmathbb_{2}(V)$, that is,
\begin{equation}
(\Tmathbb_{1} \vee \Tmathbb_{2})(V)
:=
\{
P \cap P' \in \Sub_{\dB}(O_{\downarrow V}) \;|\;
P,\,P' \in \Tmathbb_{1}(V) \cup \Tmathbb_{2}(V)
\}.
\end{equation}
Let $\jmathbold(\Tmathbb_{j})$ be the set of all translation presheaves of a truth sheaf $\Tmathbb_{j}$.
Since the translation condition (\ref{eq:translation condition of truth sheaf}) is 
equivalent to $\Tmathbb(\flat(V)) = (\varrho_{O})_{V}^{-1}(\Tmathbb_{j}(V))$ for all $V \in \Vbold$,
$\jmathbold(\Tmathbb_{j})$ is closed for $\vee$ and $\wedge$ defined above.

Also for every truth sheaf $\Tmathbb_{j}$, 
we can define its canonical translation $\jmathboldmax(\Tmathbb_{j})$ 
that is the largest one among the translations satisfying (\ref{eq:translation condition of truth sheaf}).
It is defined as the pullback of  
$\begin{xy} {\ar @{>->}  (0,0)*{\varrho_{O}^{-1}(\Tmathbb_{j})\;};(20,0)*{\;\flat^{*}(\Pmathbb_{\dB} O) }}  \end{xy}$ 
along the morphism $\Pmathbb_{\dB} O \xrightarrow{\zeta_{\Pmathbb_{\dB} O}} \flat^{*}(\Pmathbb_{\dB} O)$:
%
\begin{eqnarray}
\jmathboldmax (\Tmathbb_{j})(V) 
& = &
\{ A \in \Sub_{\dB}(O_{\downarrow V}) \;|\; (\zeta_{\Pmathbb O})_{V}(A) \in \varrho_{O}^{-1}(\Tmathbb_{j})(V)\} \nonumber\\
& = &
\{ A \in \Sub_{\dB}(O_{\downarrow V}) \;|\; A_{\downarrow \flat(V)}\in \varrho_{O}^{-1}(\Tmathbb_{j})(V)\} \nonumber\\
& = &
\{ A \in \Sub_{\dB}(O_{\downarrow V}) \;|\; \flat^{*}(A_{\downarrow \flat(V)}) \in \Tmathbb_{j}(\flat(V))\} \nonumber\\
& = &
\{ A \in \Sub_{\dB}(O_{\downarrow V}) \;|\; \flat^{*}(A_{\downarrow V}) \in \Tmathbb_{j}(V)\} .
\end{eqnarray}

Clearly, if $\Tmathbb_{j}$ is a truth sheaf, every $\jmathboldmax (\Tmathbb_{j})(V)$ is a filter,
hence, $\jmathboldmax (\Tmathbb_{j})$ a truth presheaf.

Next, let us define, for each $V \in \Vbold$ and $W \in \Ucalflat(V)$, $\Rmathbb_{V;\, W} \subseteq \Sub_{\dB}(O_{\downarrow V})$ by
\begin{equation}
\Rmathbb_{V;\, W}
:=
\{ A_{\downarrow V} \in \Sub_{\dB}(O_{\downarrow V}) \;|\; 
A \in (\varrho_{O}^{-1}(\Tmathbb_{j}))(W)\},
\end{equation}
and $\Rmathbb_{V} \subseteq \Sub_{\dB}(O_{\downarrow V})$ by
\begin{equation}
\Rmathbb_{V} := \bigcup \{ \Rmathbb_{V;\, W} \}_{W \in \Ucalflat(V)}.
\end{equation}
We define $\jmathboldmin(\Tmathbb_{j})$ by
\begin{equation}
\jmathboldmin(\Tmathbb_{j})(V) :=
\begin{cases}
\Fcal_{V}(\Rmathbb_{V}) & \mbox{if }\, \Ucalflat(V) \ne \emptyset \\
\emptyset & \mbox{if }\, \Ucalflat(V) = \emptyset,
\end{cases}
\end{equation}
where $\Fcal_{V}(\Rmathbb_{V})$ is the smallest filter in $\Sub_{\dB} O_{\downarrow V}$ including $\Rmathbb_{V}$.
%
%
\begin{prp}
For every $\Tmathbb_{j} \in \Tpre(O)$,
$\jmathboldmin(\Tmathbb_{j})$ is the smallest translation of $\Tmathbb_{j}$.
\end{prp}
Proof.
We prove $\jmathboldmin(\Tmathbb_{j}) $ to be a presheaf.
Suppose that $A \in \jmathboldmin(\Tmathbb_{j})(V)$.
This is equivalent to that there exists a finite subset $\Smathbb_{V}$ of $\Rmathbb_{V}$
such that $\wedge \Smathbb_{V} \subseteq A$, since $\jmathboldmin(\Tmathbb_{j})(V)$ is a filter \cite{DP90}.
Therefore, for every $V' \subseteqVbold V$, we have 
$\wedge (\Smathbb_{V})_{\downarrow V'} \subseteq A_{\downarrow V'}$,
where $ (\Smathbb_{V})_{\downarrow V'} \equiv \{ B_{\downarrow V'} \;|\; B \in \Smathbb_{V} \}$ is a finite subset of $\Rmathbb_{V'}$.
Thus, $A_{\downarrow V'} \in \jmathboldmin(\Tmathbb_{j})$.

To show that $\jmathboldmin(\Tmathbb_{j})$ is a translation of $\Tmathbb_{j}$,
note that $V \in \Ucalflat(\flat(V))$.
We have, for every $W \in \Ucalflat(\flat(V)) \setminus \{ V \}$,
\begin{equation}
\Rmathbb_{\flat(V);\, W} = 
\{A_{\downarrow \flat(V)} \;|\; A \in (\varrho_{O})^{-1}(\Tmathbb_{j})(W)\}
\subseteq 
 (\varrho_{O})^{-1}(\Tmathbb_{j})(V),
\end{equation}
whereas,
\begin{equation}
\Rmathbb_{\flat(V);\, V} = 
\{A_{\downarrow \flat(V)} \;|\; A \in (\varrho_{O})^{-1}(\Tmathbb_{j})(V)\}
=
 (\varrho_{O})^{-1}(\Tmathbb_{j})(V).
\end{equation}
Thus, we obtain
\begin{equation}
\Rmathbb_{\flat(V)} = \Rmathbb_{\flat(V);\, V} \cup (\bigcup \{\Rmathbb_{\flat(V);\, W}\}_{W \in  \Ucalflat(\flat(V)) \setminus \{ V \}}) =  (\varrho_{O})^{-1}(\Tmathbb_{j})(V),
\end{equation}
hence,
\begin{eqnarray}
\jmathboldmin(\Tmathbb_{j})(\flat(V)) 
& = &
\Fcal_{\flat(V)}( \Rmathbb_{\flat(V)}) \nonumber\\
& = &
\Fcal_{\flat(V)}( (\varrho_{O})^{-1}(\Tmathbb_{j})(V)) \nonumber\\
& = & 
(\varrho_{O})^{-1}(\Tmathbb_{j})(V),
\end{eqnarray}
where from the second line to the third, we used the fact that $ (\varrho_{O})^{-1}(\Tmathbb_{j})(V)$ itself is a filter.

Finally, we show that $\jmathboldmin(\Tmathbb_{j})$ is the smallest translation of $\Tmathbb_{j}$.
It suffices to show for $V \in \Vbold$ such that $\Ucalflat(V) \ne \emptyset$. 
Let $\Tmathbb$ be an arbitrary translation of $\Tmathbb_{j}$.
Suppose that $A \in \jmathboldmin(\Tmathbb_{j})(V)$.
Then, there exists a finite subset $\Smathbb_{V}$ of $\Rmathbb_{V}$ such that $\wedge \Smathbb_{V} \subseteq A$.
On the other hand, for every $B \in \Smathbb_{V}$, there exists a $W \in \Ucalflat(V)$ such that $B \in \Rmathbb_{V; \, W}$;
that is, there exists a $C \in (\varrho_{O})^{-1}(\Tmathbb_{j})(W) = \Tmathbb(\flat(W))$ such that $B = C_{\downarrow V}$.
This implies that $B = \Tmathbb(V \hookrightarrow \flat(W))(C) \in \Tmathbb(V)$.
Thus, $\Smathbb_{V} \subseteq \Tmathbb(V)$, hence, $\wedge \Smathbb_{V} \in \Tmathbb(V)$, which implies $A \in \Tmathbb(V)$ since $\Tmathbb(V)$ is a filter.
\qed

\begin{thm}
\label{thm:coarse-graining of truth object}
For every truth sheaf $\Tmathbb_{j}$, 
$\jmathbold(\Tmathbb_{j})$ is a lattice that is given by
\begin{equation}
\jmathbold(\Tmathbb_{j}) = 
\{ \Tmathbb \in \Tpre \;|\; 
\jmathboldmin(\Tmathbb_{j}) \subseteq \Tmathbb \subseteq \jmathboldmax(\Tmathbb_{j})
\}.
\end{equation}
\end{thm}

Every truth sheaf $\Tmathbb_{j}$ determines a lattice of truth presheaves consisting of translations $\Tmathbb_{j}$.
Not all truth presheaves, however, are not translations of truth sheaves. 
In fact, if $\Tmathbb$ is a translation of $\Tmathbb_{j}$, $(\rho_{O})_{V}((\flat^{*}\Tmathbb)(V)) = \Tmathbb_{j}(V)$ needs to be satisfied.
However, in general for such $\Tmathbb$, it only holds that $(\flat^{*}\Tmathbb)(V) \subseteq (\rho_{O})_{V}^{-1}((\rho_{O})_{V}((\flat^{*}\Tmathbb)(V)))$.
Consequently, the set $\Tpre$ of truth presheaves is divided into the pairwise disjoint lattices each of which corresponds to one and the same truth sheaf and the other truth presheaves that fail to be translations.



\appendix

\section{Presheaf-Based Truth-Value Assignment}
\label{sec:DI}

In this appendix, we give a brief explanation of the truth-value assignment method developed by D\"{o}ring and Isham \cite{DI12}, for the purpose of convenience for comparison with the present truth-value assignment on $j$-sheaves.

The main ingredient is the spectral presheaf $\Sigma$,
which is a presheaf such that, for each $V \in \Vbold$,
$\Sigma(V)$ is the Gelfand space on $V$,
and for $V' \subseteqVbold V$ and $\sigma \in \Sigma(V)$,
$\Sigma(V' \hookrightarrow V)(\sigma)$ is a restriction of $\sigma$ to $V'$.  
The spectral presheaf plays a role of state space;
every proposition on a given quantum system is assumed to be representable as a clopen subobject $S$ of the spectral presheaf $\Sigma$,
where $S$ is called a clopen subobject of $\Sigma$ when $S(V)$ is  a closed and open subset of $\Sigma(V)$.
Thus, the collection $\Sub_{\cl}(\Sigma)$ of all clopen subobjects of $\Sigma$ can be regarded as a space of propositions.
It is internalized to $\Vboldhat$ by the clopen power object $\Pmathbb_{\cl} \Sigma \equiv \Omega^{\Sigma}$ of $\Sigma$,
which is expressed as
\begin{equation}
(\Pmathbb_{\cl} \Sigma)(V) := \Sub_{\cl} (\Sigma_{\downarrow V}),
\end{equation}
and 
\begin{equation}
(\Pmathbb_{\cl} \Sigma)(V' \hookrightarrow V):\Sub_{\cl} (\Sigma_{\downarrow V}) \to \Sub_{\cl} (\Sigma_{\downarrow V'});
S \mapsto S_{\downarrow V}.
\end{equation}
There is a bijection from $\Sub_{\cl}(\Sigma)$ to $\Gamma(\Pmathbb_{\cl} \Sigma) := \Hom(1,\Pmathbb_{\cl} \Sigma)$ which assigns to each proposition $P$ its name $\lceil P \rceil$ defined by
\begin{equation}
\lceil P \rceil_{V} := P_{\downarrow V}.
\end{equation}
Here, for each presheaf $Q \in \Vboldhat$ and $V \in \Vbold$,
we define $Q_{\downarrow V} \in \Vboldhat$ as the downward restriction of $Q$
to $V' \subseteqVbold V$:
\begin{equation}
Q_{\downarrow V}(V') := 
\begin{cases}
Q(V') & \mbox{if }\, V' \subseteqVbold V, \\
\emptyset & \mbox{otherwise} .
\end{cases}
\label{eq:downarrow 1}
\end{equation}
and for each $V'' \subseteqVbold V'$,
\begin{equation}
Q_{\downarrow V}(V'' \hookrightarrow V') := 
\begin{cases}
Q(V'' \hookrightarrow V') & \mbox{if }\, V' \subseteqVbold V ,\\
\emptyset \xrightarrow{!} Q(V'') & \mbox{otherwise} .
\end{cases}
\label{eq:downarrow 2}
\end{equation}

D\"{o}ring and Isham gave other ways to express propositions.
They are based on the outer presheaf $O$ that is defined by
\begin{equation}
O(V):=\Pcal(V)
\label{eq:outer presheaf 1}
\end{equation} 
and for $V' \subseteqVbold V$,
\begin{equation}
O(V' \hookrightarrow V):O(V) \to O(V'); \Phat \mapsto \delta (\Phat)_{V'}.
\label{eq:outer presheaf 2}
\end{equation}
Here, $\Pcal(V)$ is the set of all projection operators in $V$
and $\delta$ the daseinization operator,
which assigns to each projection operator $\Phat$ a collection $\delta(\Phat) := \{ \delta(\Phat)_{V} \}_{V \in \Vboldhat}$,
each element $\delta(\Phat)_{V}$ of which is defined by
\begin{equation}
\delta(\Phat)_{V} := \bigwedge \{\alphahat \in \Pcal(V) \;|\; \Phat \preceq \alpha  \}.
\label{eq:daseinization}
\end{equation}
Obviously, $\delta(\Phat)$ is a global element of the outer presheaf $O$.
Note that for every $V' \subseteqVbold V$, it follows 
\begin{equation}
\delta(\delta(\Phat)_{V})_{V'} = \delta(\Phat)_{V'}.
\end{equation}
This equality is often used in the text.

D\"{o}ring \& Isham proved that
\begin{equation}
\Sub_{\dB} (O) \simeq \Hyp (O) \simeq \Sub_{{\rm cl}}(\Sigma),
\label{eq:SubHypSub}
\end{equation}
and hence for every $V \in \Vbold$,
\begin{equation}
\Sub_{\dB} (O_{\downarrow V}) \simeq \Hyp (O_{\downarrow V}) \simeq \Sub_{{\rm cl}}(\Sigma_{\downarrow V}).
\label{eq:SubVHypVSubV}
\end{equation}
Here, $\Sub_{\dB} (O) $ is the collection of subobjects $B \subseteq O$
such that, for every $V \in \Vbold$,
$B(V) \subseteq O(V)$ is a downward closed set of $O(V)$ with a top element.
Obviously, it is a complete Boolean lattice.
On the other hand, $\Hyp(O)$ is the collection of all hyper-elements of $O$,
where a hyper-element $h$ of $O$ is a collection $\{ \hhat_{V} \in O(V) \}_{ V \in \Vbold }$
that satisfies, for every $V' \subseteqVbold V$, 
\begin{equation}
O(V' \hookrightarrow V)(\hhat_{V}) = \delta(\hhat_{V})_{V'} \preceq \hhat_{V'}.
\label{eq:hyper-element of outer presheaf}
\end{equation}
The bijection relation (\ref{eq:SubHypSub}) is given by the function $k:\Hyp(O) \xrightarrow{\sim} \Sub_{\cl}(\Sigma)$ defined by
\begin{equation}
(k(h))(V):=\alpha_{V}(\hhat_{V}),
\end{equation}
and $c:\Sub_{\dB}(O) \xrightarrow{\sim} \Hyp(O)$ defined by
\begin{equation}
c(A)_{V} := \vee A(V).
\end{equation}
Here, the function $\alpha_{V}:\Pcal(V) \to \Cl(\Sigma(V))$, where $\Cl(\Sigma(V))$ is the collection of all clopen subsets of $\Sigma(V)$, is defined as
\begin{equation}
\alpha_{V}(\Phat) := \{ \sigma \in \Sigma(V) \;|\; \sigma(\Phat) = 1 \}.
\label{eq:alphaV}
\end{equation}
Bijections for (\ref{eq:SubVHypVSubV}) are given as the restrictions of $k$ and $c$ to subalgebras of $V$.

In particular, every projection $\Phat$ defines a proposition presheaf, 
the global element  $\delta(\Phat) \in \Gamma O\subseteq \Hyp(O)$.
That is, as an element of $\Sub_{\cl} \Sigma$, $\delta(\Phat)$ is given by
\begin{eqnarray}
(\delta(\Phat))(V)& := & 
\alpha_{V}(\delta(\Phat)_{V}) \nonumber\\
& = &
 \{ \sigma \in \Sigma(V) \;|\;\sigma(\delta(\Phat)_{V})) = 1 \},
\end{eqnarray}
and as that of $\Sub_{\dB}(O)$,
\begin{eqnarray}
(\delta(\Phat))(V)& := & 
c_{V}^{-1}(\delta(\Phat)_{V}) \nonumber\\
& = &
\{\ahat \in O(V) \;|\; \ahat \preceq \delta(\Phat)_{V} \}.
\end{eqnarray}

Each proposition $P \in \Sub_{\cl}(\Sigma)$ is assigned a truth value relative to a truth object $\Tmathbb$,
a subobject of $\Pmathbb_{\cl} \Sigma$ (or, equivalently that of $\Pmathbb_{\dB} O$)
of which global elements give truth propositions.
Let $\tau$ be the characteristic morphism of $\Tmathbb$;
That is, the diagram
\begin{equation}
\xymatrix{
 \Tmathbb \ar [rr] ^{!} \ar @{>->} [dd] && 1 \ar @{>->} [dd] ^{\true} \\
&& \\
 \Pmathbb_{\cl} \Sigma \ar [rr] _{\tau} && \Omega \\
}
\label{eq:truth-value pullback}
\end{equation}
is a pullback.
Then, for each proposition $P$, 
its truth-value $\nu(P;\Tmathbb) \in \Gamma\Omega$ is given by
\begin{equation}
\nu(P;\Tmathbb) = \tau \circ \lceil P \rceil,
\end{equation}
or more precisely,
\begin{equation}
\nu(P;\Tmathbb)_{V} 
=
\{V' \subseteqVbold V \;|\; P_{\downarrow V} \in \Tmathbb(V') \}.
\end{equation}

In \cite{DI12}, D\"{o}ring and Isham defined generalized truth object
\begin{equation}
\Tmathbbrhor(V)
:= \{
A \in \Sub_{\dB}(O_{\downarrow V}) \;|\;
\forall V' \subseteqVbold V \;(\tr(\rho (\vee A(V'))) \ge r)
\},
\label{eq:truth presheaf rho r}
\end{equation}
which gives propositions that are true at least a probability $r \in [0,1]$ in a mixed state expressed by a density matrix $\rho$.
Under the truth presheaf $\Tmathbbrhor$,
the truth-value of $\delta(\Phat)$ is evaluated as
\begin{eqnarray}
\nu_{j}(\delta_{j}(\Phat);\Tmathbbrhor)_{V}
& = &
\{ V' \subseteqVbold V \;|\;
\forall V'' \subseteqVbold V' \; (\tr (\rho (\delta(\Phat)_{V''}) )\ge r)
\} \nonumber\\
& = &
\{ V' \subseteqVbold V \;|\;
\tr (\rho (\delta(\Phat)_{V'}) )\ge r
\}.
\end{eqnarray}
If we take $\rho = \varphipro$ and $r=1$, 
the truth presheaf $\Tmathbbphi := \Tmathbb^{\varphipro, \, 1}$ and the truth-value of $\delta(\Phat)$ are given by
\begin{equation}
\Tmathbbphi(V) := \{ A \in \Pmathbb_{\dB} O (V) \;|\; \forall V' \subseteqVbold V \; (\delta(\varphipro)_{V'} \in A(V')) \},
\label{eq:truth presheaf for vector state}
\end{equation}
and
\begin{eqnarray}
\nu(\delta(\Phat);\Tmathbbphi) (V) 
& = &
\{ V' \subseteqVbold V \;|\; \delta(\Phat)_{V'} \in \Tmathbbphi (V')\} \nonumber\\
& = &
\{V' \subseteqVbold V \;|\; \forall V'' \subseteqVbold V' \;(\delta(\varphipro)_{V''} \in \delta(\Phat)(V'')) \} \nonumber\\
& = &
\{V' \subseteqVbold V \;|\; \forall V'' \subseteqVbold V' \;(\delta(\varphipro)_{V''} \preceq \delta(\Phat)_{V''}) \} \nonumber\\
& = &
\{V' \subseteqVbold V \;|\; \forall V'' \subseteqVbold V' \;(\varphipro\preceq \delta(\Phat)_{V''}) \} \nonumber\\
& = &
\{ V' \subseteqVbold V \;|\; \varphipro \preceq \delta(\Phat)_{V'}  \},
\label{eq:nu 2}
\end{eqnarray}
respectively.

\section{Mathematical Miscellany}
\label{sec:misc}

In the text, some propoerties of the functors $\flat: \Vbold \to \Vbold$ and $\flat^{*}:\Vboldhat \to \Vboldhat$ are used. 
In this appendix, we explain them for convenience.

Throughout the text, we use the relation
$
\flat \flat = \flat
$
without notice.
Furthermore, the following fact is often used: 
for any presheaf $Q \in \Vboldhat$
and any subsheaf $A$ of $\flat^{*} Q$,
\begin{equation}
A = \flat^{*} A.
\end{equation}
To show this, let $A \in \Vboldhat$ be a subobject of $\flat^{*} Q$.
The closure $\bar{A}$ of $A$ in $\flat^{*}Q$ is given by
\begin{eqnarray}
\bar{A} (V) 
& = &
\{  q \in (\flat^{*}Q)(V) \;|\; (\flat^{*}Q)(\flat(V)\hookrightarrow V)(q) \in A(\flat(V)) \} \nonumber \\
& = &
\{ q \in Q(\flat((V))) \; | \; q \in A(\flat(V)) \} \nonumber \\
& = &
A(\flat(V)) \nonumber \\
& = &
\flat^{*} A (V).
\end{eqnarray}
Thus, $A$ is closed (hence, a sheaf) if and only if $A = \flat^{*} A$.

When we deal with power objects, the following relations are crucially important:
for each $Q \in \Vboldhat$ and $V \in \Vbold$, we have
\begin{equation}
\flat^{*}(Q_{\downarrow V}) = \flat^{*}(Q_{\downarrow \flat(V)}),
\label{eq:flat1}
\end{equation}
\begin{equation}
\flat^{*}((\flat^{*}Q)_{\downarrow V}) =
\flat^{*}(Q_{\downarrow V}),
\label{eq:flat3}
\end{equation}
and furthermore, for any $V' \subseteqVbold V$,
\begin{equation}
\flat^{*}((\flat^{*}(Q_{\downarrow V}))_{\downarrow V'}) =
\flat^{*}(Q_{\downarrow V'}).
\label{eq:flat2}
\end{equation}
They can be proved straightforwardly.

In section \ref{sec:translation rule}, we treat a relation between $\flat^{*}\Pmathbb$ and $\Pmathbb_{j} \flat^{*}$.
They are functors from $\Vboldhat$ to $\Sh_{j}\Vboldhat$,
and there exists a canonical natural transformation $\flat^{*}\Pmathbb \xrightarrow{\varrho} \Pmathbb_{j}\flat^{*}$,
which is defined as follows.
First, note that, for each presheaf $Q \in \Vboldhat$ and $V \in \Vbold$,
\begin{equation}
\flat^{*}(\Pmathbb Q)(V) =\Pmathbb Q (\flat(V)) \simeq \Hom (Q_{\downarrow \flat(V)}, \Omega),
\end{equation}
and 
\begin{equation}
\Pmathbb_{j}(\flat^{*} Q)(V) 
\simeq 
\Hom(\flat^{*}(Q_{\downarrow V}), \Omega_{j})
=
\Hom(\flat^{*}(Q_{\downarrow \flat(V)}), \Omega_{j}).
\end{equation}

Let $S$ be a subobject of $Q_{\downarrow V}$ and $Q_{\downarrow V} \xrightarrow{\chi} \Omega$ be the characteristic morphism of $S$ in $\Vboldhat$.
Then, we have the following commutative diagram:
\begin{equation}
\xymatrix{
& S \ar [rrr] ^{!} \ar @{>->} _{\iota} [dl] \ar @{.>} [dd] ^(0.3){\zeta_{S}}&&& 1\ar @{>->} [ld] ^(0,6){\true} \ar @{=} [dd] \\
Q_{\downarrow \flat(V)} \ar [dd] _{\zeta_{Q_{\downarrow \flat(V)}}} \ar [rrr] _(0.3){\chi}&&& \Omega \ar [dd] ^(0.3){\zeta_{\Omega}}& \\
& \flat^{*}S \ar @{>.>} [ld] _{\flat^{*} \iota} \ar @{.>} [rrr] ^{!} 
\ar @{.}@<0.5mm> [dd] \ar @{.}@<-0.5mm> [dd] &&& 1 \ar @{>->} [ld] ^(0.65){\flat^{*}  \true} \ar @{=} [dd] \\
\flat^{*} (Q_{\downarrow \flat(V)}) \ar [rrr] _(0.3){\flat^{*} \chi} \ar @{=} [dd] &&& \flat^{*}\Omega \ar [d] ^{\flat^{*} r } & \\
& \flat^{*}S \ar @{>.>} [ld] _{\flat^{*} \iota}  \ar @{.} [rr] ^{!} && \flat^{*}\Omega_{j} \ar @{.>} [r] \ar [d] ^{\zeta^{-1}_{\Omega_{j}}} _{\vertsim} & 1 \ar @{>->} [ld] ^{\true_{j}}\\
\flat^{*} (Q_{\downarrow V}) \ar [rrr] _{\flat^{*} (r \circ \chi)}&&& \Omega_{j}  \;. & \\
}
\label{eq:sbox}
\end{equation}
Here, the top square is a pullback, and hence, so is the bottom one,
because of the left-exactness of the associated sheaf functor $\flat^{*}$. 
Thus, we define $(\varrho_{Q})_{V}$ as a function that maps the top square to the bottom one;
that is, as a function from $\flat^{*}(\Pmathbb Q) (V)$ to $(\Pmathbb_{j} (\flat^{*}Q)) (V)$,
it is defined by
\begin{equation}
(\varrho_{Q})_{V}(S) := \flat^{*} S ,
\end{equation}
and hence, as a function from $\Hom (Q_{\downarrow \flat(V)}, \Omega)$ to $\Hom(\flat^{*}(Q_{\downarrow \flat(V)}), \Omega_{j})$,
\begin{equation}
(\varrho_{Q})_{V}(\chi) := \flat^{*}(r \circ \chi).
\end{equation}
We can straightforwardly show that $(\varrho_{Q})(V)$ is natural with respect to 
$Q \in \Vboldhat$ and $V \in \Vbold$.

In the text, the case where $Q$ is the outer presheaf $O$ is treated.
As easily shown, the restriction of $\varrho_{O}$ to $\flat^{*}(\Pmathbb_{\dB}O)$ takes values on $\Pmathbb_{j \, \dB}(\flat^{*} O)$ and the diagram
\begin{equation}
\xymatrix{
\flat^{*}(\Pmathbb_{\dB}O) \ar [rr] ^{\varrho_{O}|_{\flat^{*}(\Pmathbb_{\dB}O)}} \ar @{>->}  [dd] &&  \Pmathbb_{j \, \dB}(\flat^{*} O)  \ar @{>->}  [dd]   \\
&& \\
\flat^{*}(\Pmathbb Q) \ar [rr] _{\varrho_{O}}&& \Pmathbb_{j} (\flat^{*}Q) \\     
}
\end{equation}
commutes.
Because of this, in section \ref{sec:translation rule}, we write $\varrho_{O}$ for the restriction $\varrho_{O}|_{\flat^{*}(\Pmathbb_{\dB}O)}$ described above.


\end{document}